%% file: grb090727.tex
\documentclass[numberedappendix,twocolappendix]{emulateapj}
\usepackage{amsmath, amsfonts, amssymb}
\usepackage{times}

\begin{document}
\title{GRB~090727 and gamma-ray bursts with early time optical emission}

\author{D.~Kopa\v c\altaffilmark{1},
S.~Kobayashi\altaffilmark{2},
A.~Gomboc\altaffilmark{1,3},
J.~Japelj\altaffilmark{1},
C.~G.~Mundell\altaffilmark{2},
C.~Guidorzi\altaffilmark{4},
A.~Melandri\altaffilmark{5},
D.~Bersier\altaffilmark{2},
Z.~Cano\altaffilmark{2},
R.~J.~Smith\altaffilmark{2},
I.~A.~Steele\altaffilmark{2},
F.~J.~Virgili\altaffilmark{2}}

\altaffiltext{1}{Department of Physics, Faculty of Mathematics and Physics, University of Ljubljana, Jadranska 19, SI-1000 Ljubljana, Slovenia.}
\altaffiltext{2}{Astrophysics Research Institute, Liverpool John Moores University, Twelve Quays House, Egerton Wharf, Birkenhead, CH41 1LD, UK}
\altaffiltext{3}{Centre of Excellence SPACE-SI, A\v sker\v ceva cesta 12, SI-1000 Ljubljana, Slovenia.}
\altaffiltext{4}{Physics Departments, University of Ferrara, via Saragat 1, I-44122, Ferrara, Italy.}
\altaffiltext{5}{INAF -- Osservatorio Astronomico di Brera, via E. Bianchi 46, 23807 Merate (LC), Italy.}
\email{drejc.kopac@fmf.uni-lj.si}

\begin{abstract}

\par We present a multi-wavelength analysis of gamma-ray burst GRB~090727, for which optical emission was detected during the prompt gamma-ray emission by the 2-m autonomous robotic Liverpool Telescope and subsequently monitored for a further two days with the Liverpool and Faulkes telescopes. Within the context of the standard fireball model, we rule out a reverse shock origin for the early time optical emission in GRB~090727 and instead conclude that the early time optical flash likely corresponds to emission from an internal dissipation processes. Putting GRB~090727 into a broader observational and theoretical context, we build a sample of $36$ gamma-ray bursts (GRBs) with contemporaneous early time optical and gamma-ray detections. From these GRBs, we extract a sub-sample of $18$ GRBs, which show optical peaks during prompt gamma-ray emission, and perform detailed temporal and spectral analysis in gamma-ray, X-ray, and optical bands. We find that in most cases early time optical emission shows sharp and steep behaviour, and notice a rich diversity of spectral properties. Using a simple internal shock dissipation model, we show that the emission during prompt GRB phase can occur at very different frequencies via synchrotron radiation. Based on the results obtained from observations and simulation, we conclude that the standard external shock interpretation for early time optical emission is disfavoured in most cases due to sharp peaks ($\Delta t/t < 1$) and steep rise/decay indices, and that internal dissipation can explain the properties of GRBs with optical peaks during gamma-ray emission.
\end{abstract}


\maketitle

\section{Introduction}
\par Gamma Ray Bursts (GRB), as their name suggests, were discovered through detection of their high-energy prompt gamma-ray emission \citep{klebesadel1973}. Subsequent detection of nearly $3000$ GRBs with the BATSE gamma-ray detector onboard the Compton Gamma Ray Observatory \citep{paciesas1999}, showed them to display a rich variety of temporal profiles in their gamma-ray light curves with single or multiple peaks. The \textit{BeppoSAX} satellite \citep{boella1997} made the first discovery of an X-ray afterglow in GRB~970228 with an accurate localisation \citep{costa1997} that led to the discovery of an associated optical counterpart several hours after the burst and, thus, GRB redshift \citep{vanparadijs1997}. 

\par Despite the growing number of gamma-ray light curves obtained to date, the rich variety in temporal and spectral properties coupled with the challenge in obtaining panchromatic observations simultaneously with prompt gamma-ray emission has limited understanding of the underlying physical processes. Although the variability in GRB light curves suggests internal dissipation processes for prompt gamma-ray emission, fundamental theoretical questions remain on the actual dissipation mechanism, the presence of thermal components and the role of magnetic fields \citep[e.g.,][]{atteiaboer2011, gomboc2012, harrison2013}.

\par The first detection of optical emission during the prompt gamma-ray phase was made for GRB~990123 \citep{akerlof1999}, which stimulated research efforts on theoretical and observational side. The launch of NASA's \textit{Swift} satellite \citep{gehrels2004} revolutionised the study of GRBs and opened a new window on the very early time properties of GRBs; in particular, the real-time dissemination of accurate ($3\,\mathrm{arcmin}$ error circles) gamma-ray positions from its Burst Alert Telescope (BAT) via the Gamma-ray Coordinates Network (GCN) \citep{barthelmy1994}, are key for driving rapid-response ground-based observations from fully autonomous robotic optical/IR telescopes such as the 2-m Liverpool Telescope -- LT \citep{steele2004, guidorzi2006}. In parallel, Swift's onboard rapid-response X-ray (XRT; \citealt{burrows2005}) and optical (UVOT; \citealt{roming2005}) telescopes provide additional temporal and spectral coverage of early time GRB properties.

\par Theoretically, several mechanisms have been proposed to predict and explain the origin of early time optical emission. Some focus on the standard fireball model (see for example the review by \citealt{piran2004}), interpreting early time optical emission as the onset of the external shock afterglow emission, either from forward-external or reverse-external shock \citep{saripiran1999, meszarosrees1999}. Reverse-external shock emission, in particular, was expected to manifest as a bright optical flash in the early time light curve \citep{kobayashi2000}. Other models explain early time optical emission as a low-energy tail of high-energy emission, produced in the internal shocks \citep{katz1994, meszarosrees1999, wei2007}. \citet{panaitescu2008} proposed that the synchrotron self-Compton (SSC) mechanism could explain gamma-ray emission in GRBs with prompt optical detection, where optical synchrotron photons act as a seed, but this model was later brought into question \citep{zou2009,piran2009}, especially by \textit{Fermi} LAT observations \citep{beniamini2011}. Similarly, \citet{zhao2010} studied the origin of prompt optical and gamma-ray emission in the context of inverse Compton, second inverse Compton and SSC models, but these mechanisms are difficult to probe due to the paucity of the \textit{Fermi} LAT detections at high energies and only a handful of corresponding bright early time optical flashes detected. Fast decay in early time optical emission could also be explained by the large-angle emission \citep{kumar2008}. Alternative mechanisms focus on Poynting flux dominated outflows \citep[e.g.,][]{kumar2007} or magnetic reconnection \citep[e.g.,][]{thompson1994, giannios2008, zhang2011}; in such cases the early time polarisation measurements \citep[e.g.,][]{mundell2007a,steele2009} could provide more details about the origin and geometry of magnetic fields. To date it remains unclear, what are the true mechanisms for prompt GRB emission and which processes cause optical emission at early times.

\par Obtaining multi-wavelength observations as quickly as possible after the initial GRB trigger is therefore crucial to provide constraints on current and future GRB emission models. Although technically challenging, a growing number of robotic telescopes are beginning to catch optical emission during prompt gamma-ray phase. In order to detect optical emission simultaneously with gamma-ray emission, the latter must last longer than the response time of a telescope, usually several tens of seconds or minutes. A combination of sustained GRB gamma-ray emission and rapid optical follow-up is therefore providing a slowly growing sample of GRBs with prompt optical-gamma-ray observations.

\par In this paper, we present a detailed multi-wavelength analysis of GRB~090727, which was detected by the \textit{Swift} satellite and promptly observed automatically by the Liverpool Telescope through the prompt and afterglow phases (Sections \ref{sect:090727}, \ref{sect:datared} and \ref{sect:dataanal}). We then compile a sample of GRBs observed to date with contemporaneous gamma-ray and optical detection (Section~\ref{sect:sample}). From this master sample, we select a sub-sample of $18$ GRBs suitable for detailed temporal analysis. We present and discuss the results, interpretations and possible scenarios for the origin of the emission in this sub-sample (Section~\ref{sect:sampledisc}), and test this sub-sample against a simple internal shock dissipation model (Section~\ref{sect:simulation}). Summary and conclusions are given in Section~\ref{sect:conclusion}.

\par Throughout the text we use the convention $F_\nu (t) \propto t^{-\alpha}\,\nu^{-\beta}$ to describe the flux density, and assume a standard cosmology with parameters: $H_0 = 71\,\mathrm{km\,s^{-1}\,Mpc^{-1}}$, $\Omega _\Lambda = 0.73$, $\Omega _\mathrm{M} = 0.27$. Best fit parameters are given at $1\sigma$ confidence level, except when stated otherwise. Times are given with respect to GRB trigger time.

\bigskip

\section{GRB~090727 observations}
\label{sect:090727}

\par
\begin{figure}[!h]
\begin{center}
\includegraphics[width=1\linewidth]{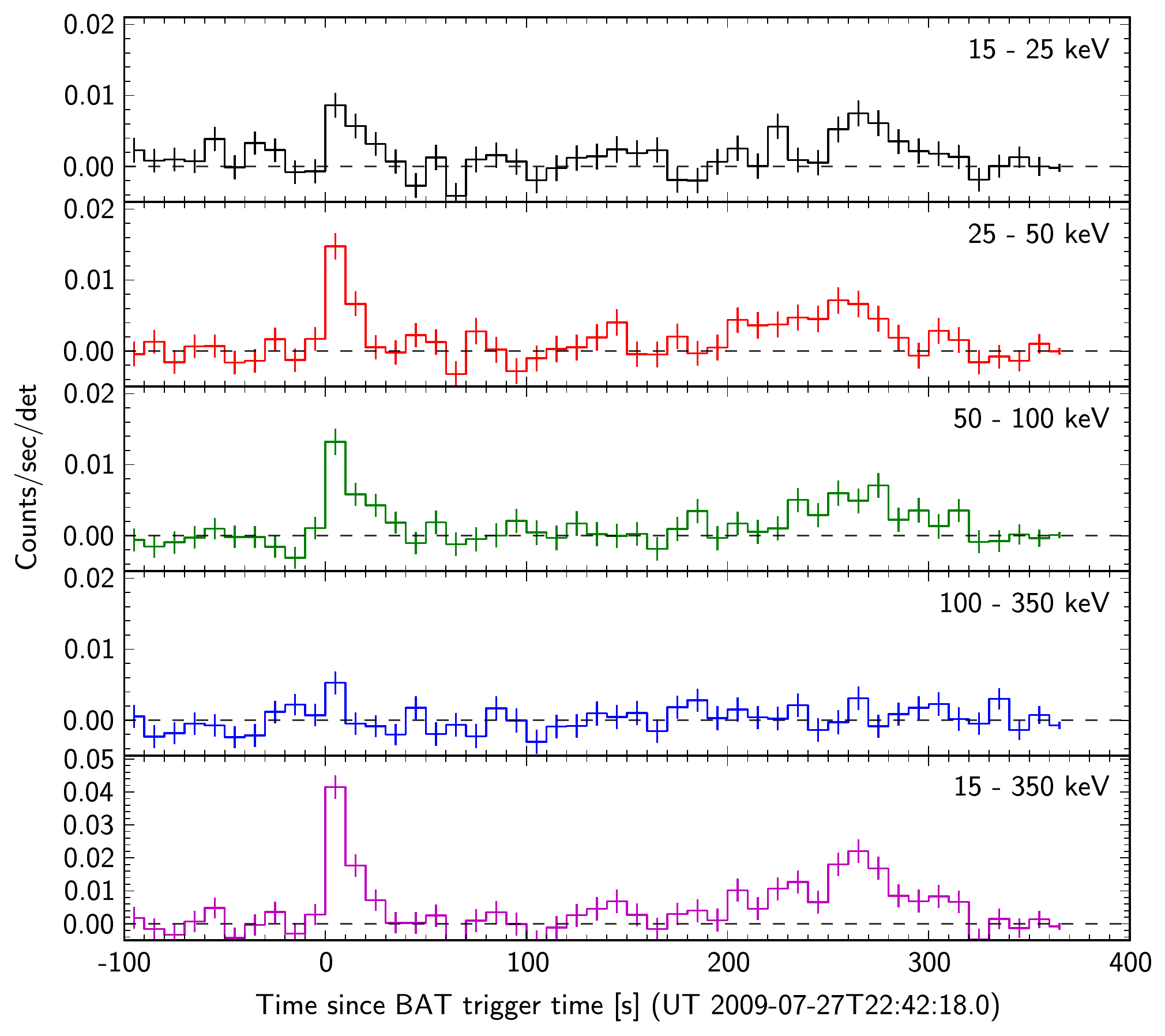} 
\caption{\label{fig:batlc} GRB~090727 gamma-ray emission light curves ($10\,\mathrm{s}$ binning) in different energy channels. There are two separate emission episodes (see the text). The second one that lasts from $t \sim 180\,\mathrm{s}$ to $t \sim 320\,\mathrm{s}$ after the trigger is softer and has a longer duration.}
\end{center}
\end{figure}

 On 27th July, 2009, at $t_0$ = 22:42:18 UT, the Burst Alert Telescope (BAT) on-board \textit{Swift} was triggered by the long GRB~090727 ($\mathrm{T}_{90} = 302 \pm 23\,\mathrm{s}$) \citep{evans2009,markwardt2009}. The gamma-ray light curve (LC) consists of two separate peaks: one narrow and intense which lasts from $t \approx 0\,\mathrm{s}$ to $t \approx 30\,\mathrm{s}$ after the trigger, and one broad and soft which lasts from $t \approx 180\,\mathrm{s}$ to $t \approx 320\,\mathrm{s}$ (Figure \ref{fig:batlc}). Gamma-ray emission detected with BAT lasted until $t \approx 320\,\mathrm{s}$.

\par
\begin{figure*}[!ht]
\begin{center}
\includegraphics[width=0.8\linewidth]{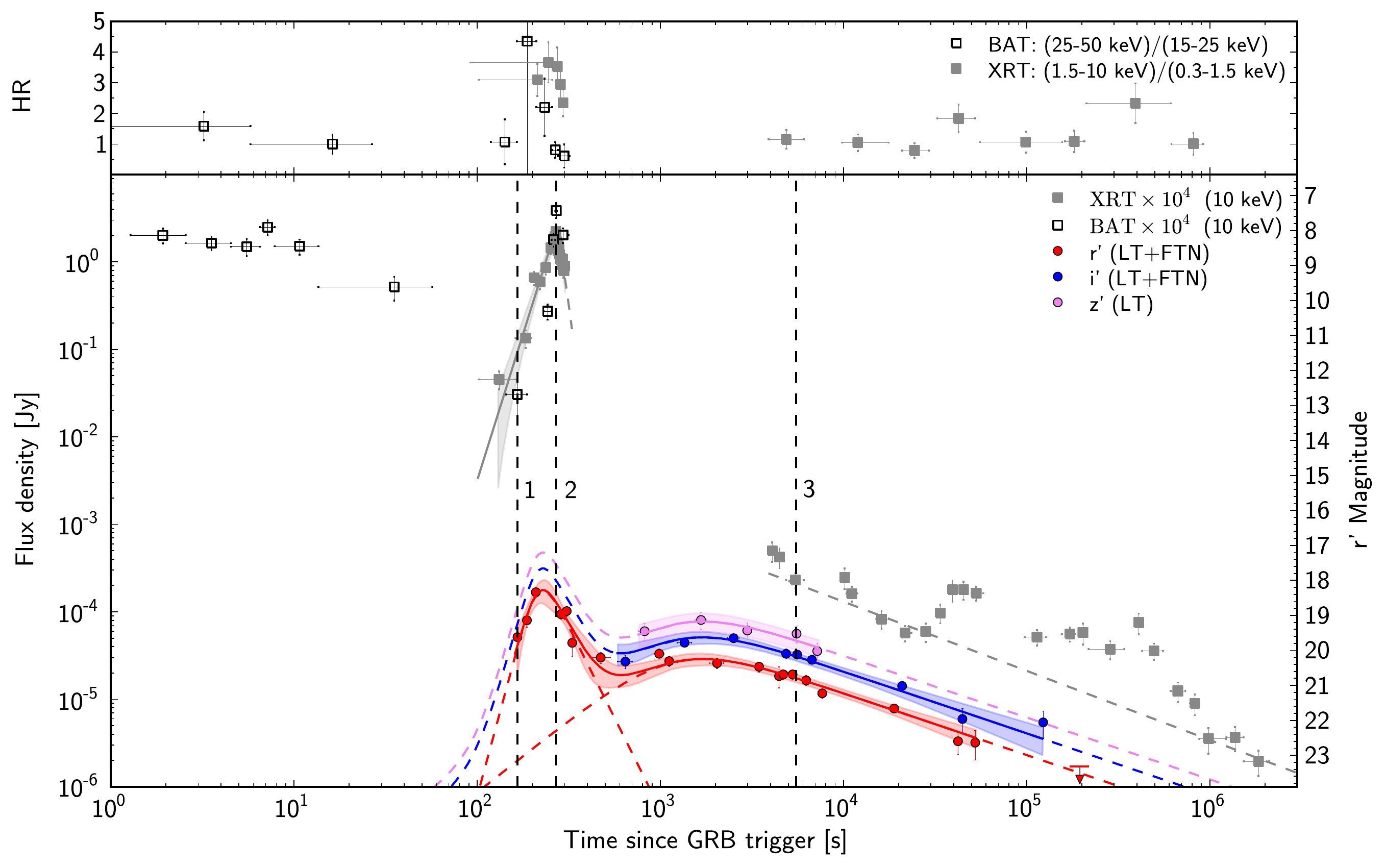} 
\caption{\label{fig:lc} GRB~090727 light curves in gamma-ray, X-ray, and optical bands, together with the hardness ratio in gamma-ray and X-ray bands (BAT and XRT data from the Swift Burst Analyser, \citealt{evans2010}). Vertical dashed lines indicate the times used to construct SEDs in Figure \ref{fig:sed}. Optical data from the Liverpool Telescope and the Faulkes telescope North were modelled using the sum of two smoothly joint broken power-law functions, as explained in the text. Early time X-ray peak was modelled using Eq. (\ref{eq:beuermann}), while late time X-ray data were modelled using a simple power-law function. Solid and dashed lines represent the best fits (Table \ref{tab:fit}), and shaded regions represent the corresponding $95\%$ confidence intervals.}
\end{center}
\end{figure*}

 Soon after the trigger, \textit{Swift} slewed to the position of the burst and the XRT started collecting data at $t = 131\,\mathrm{s}$, while gamma-ray emission was still ongoing. The X-ray light curve shows peak at $t \approx 267\,\mathrm{s}$, similar to the last gamma-ray peak (Figure \ref{fig:lc}). Between $t \approx 300\,\mathrm{s}$ and $t \approx 4000\,\mathrm{s}$ XRT did not collect any data due to the Earth limb constraint. After that, the XRT light curve shows a power-law decay with several non-prominent flares. At $t \gtrsim 2.1\times 10^6\,\mathrm{s}$ X-ray emission is no longer detected.

\par The Liverpool Telescope (LT) responded automatically \citep{guidorzi2006} to the \textit{Swift} BAT trigger, slewing to the GRB position at RA(J2000) = 21$^\mathrm{h}$03$^\mathrm{m}$50{\hbox{$.\!\!^{\rm s}$}6, Dec. (J2000) = +64$^\circ$55$^\prime$30$^{\prime \prime}$ to begin automatic optical follow-up observations at $t = 166\,\mathrm{s}$ in \textit{r'}, \textit{i'} and \textit{z'} SDSS filters \citep{smithmundell2009}. This resulted in simultaneously sampled gamma-ray, X-ray, and optical LCs around $t \approx 200-300\,\mathrm{s}$, when peak occurred in all energy bands (Figure \ref{fig:lc}). Optical follow-up observations with the LT continued until $t \approx 5.5$ hours after the trigger. Later we manually triggered the Faulkes Telescope North (FTN) to obtain additional data at $t \approx 14$ hours and $t \approx 34$ hours in the \textit{R} and \textit{i'} filters. Two days after the burst we manually triggered the LT for some deep observations, but the optical afterglow was no longer detected to limiting magnitude $\mathrm{\textit{r'}} > 23.3$.

\par Several other optical telescopes also observed the afterglow of GRB~090727. Among them, CAHA \citep{gorosabel2009} and SAO RAS \citep{moskvitin2009} detected the afterglow at $\approx 45$ minutes after the trigger. UVOT \citep{chester2009} and GRT \citep{sakamoto2009} did not detect the afterglow and thus only provided the upper limits. No redshift was measured for this GRB.

\section{GRB~090727 data reduction}
\label{sect:datared}

\subsection{Gamma-ray and X-ray data}
\par We used BAT and XRT data from the Swift Burst Analyser \citep{evans2010} to plot the LCs in Figure \ref{fig:lc}. We then analysed gamma-ray and X-ray spectra in different time epochs. For gamma-ray spectra we used data from the \textit{Swift} archive and \texttt{batbinevt} tool to create spectra files at specific time intervals (we further used \texttt{batupdatephakw} and \texttt{batphasyserr} to apply corrections to the spectra). For X-ray spectra we used the ``Create time-sliced spectra'' option on the XRT spectrum repository \citep{evans2009b} to create spectra files at specific time intervals, and \texttt{XSPECv12.6} software \citep{arnaud1996} to analyse the spectra. 

\par To fit the gamma-ray spectra we used a power-law model in the $15-150\,\mathrm{keV}$ energy range. For the X-ray spectrum, we first grouped the channels to have minimum $20$ counts in each bin. We used the combination of power-law and photoelectric absorptions (wabs$^\ast$zwabs) in $0.3-10\,\mathrm{keV}$ energy range. We fixed one contribution of photoelectric absorption to the value of the Galactic equivalent X-ray column density along the direction to GRB: $N_\mathrm{HI}^\mathrm{Gal.} = 2.5\times 10^{21} \, \mathrm{cm^{-2}}$ \citep{kalberla2005}. The second contribution (at $z=0$) was determined by fitting late time spectrum from $t = 4\times 10^3$ on, when hardness ratio becomes constant within the errors. We found the value of $N_\mathrm{HI} (z) > 0.6\times 10^{21} \, \mathrm{cm^{-2}}$, in excess of the Galactic value. This value is a lower limit because the (unknown) redshift was fixed to $z=0$ for fitting. Fit results are presented in Table \ref{tab:batxrt}. 

\par Our analysis of the gamma-ray spectrum shows that hardness ratio (HR) and photon index ($\Gamma$) are consistent within the margin of error throughout the overall event (Figure \ref{fig:lc} and Table \ref{tab:batxrt}). These results are also consistent with the results from the time-averaged spectrum reported in the Swift-BAT refined analysis GCN circular \citep{markwardt2009} and in the GCN Report 243.1. The X-ray spectrum shows change both in HR and photon index. During the X-ray peak at $t \approx 267\,\mathrm{s}$, the photon index is lower than at later times (Table \ref{tab:batxrt}). The early time X-ray photon index is thus consistent, within the margin of error, with the gamma-ray photon index. We also identified a late time peak in the X-ray LC at $t \approx 4\times 10^4\,\mathrm{s}$, but a fit to the spectrum using only data during this peak shows no difference compared to the fit when using all data from $t = 4000\,\mathrm{s}$ onwards.

\begin{deluxetable}{cccc}[!h]
\tablecaption{GRB~090727: Gamma-ray and X-ray spectral indices.}
\tabletypesize{\footnotesize}
\tablewidth{\linewidth}
\tablehead{
\colhead{Data} &
\colhead{Interval} &
\colhead{Photon index} & 
\colhead{$\chi ^2 _\mathrm{red}$ (d.o.f.)} \\
\colhead{} &
\colhead{$t\,[\mathrm{s}]$} &
\colhead{$\Gamma$} & 
\colhead{}
}
\startdata
BAT: & & & \\
Total & $0$ - $319$ & $1.34 ^{+0.26}_{-0.26}$ & $1.0$ ($56$) \\
Epoch 1 & $0$ - $200$ & $1.23 ^{+0.46}_{-0.46}$ & $1.1$ ($56$) \\
Epoch 2 & $260$ - $300$ & $1.46 ^{+0.49}_{-0.46}$ & $1.2$ ($56$) \\
 & & & \\
XRT: & & & \\
Epoch 1 (PC) & $101$ - $257$ & $1.29 ^{+0.25}_{-0.25}$ & $0.9$ ($7$) \\
Epoch 2 (WT) & $260$ - $300$ & $1.38 ^{+0.18}_{-0.17}$ & $0.8$ ($34$) \\
Epoch 3 (PC) & $4\times 10^3$ - $2\times 10^6$ & $2.31 ^{+0.29}_{-0.24}$ & $1.0$ ($26$) \\
 & & & \\
BAT + XRT: & & & \\
Epoch 2 & $260$ - $300$ & $1.51 ^{+0.08}_{-0.07}$ & $1.0$ ($95$)
\enddata
\tablecomments{BAT data were fitted with the power-law model in $15-150\,\mathrm{keV}$ band. XRT data from Photon Counting (PC) or Windowed Timing (WT) mode have been fitted with the power-law and the photoelectric absorption (wabs$^\ast$zwabs) models in $0.3-10\,\mathrm{keV}$ band. Time intervals are given with respect to the GRB trigger time $t_0$. The confidence intervals are at $90\%$.}
\label{tab:batxrt}
\end{deluxetable}

\subsection{Optical data}
\par LT and FTN optical data were calibrated against $10$ non-saturated USNO-B1 stars in the field. We converted the USNO \textit{R2} and \textit{I} catalogued magnitudes to \textit{r'i'z'} magnitudes using \citet{jordi2006}. After photometric calibration we first corrected the magnitudes for a relatively high Galactic extinction of $A_\mathrm{\textit{V}} = 1.64\,\mathrm{mag}$ \citep{schlegel1998}. Specifically, for each filter using Galactic extinction profile \citep{cardelli1989}, the extinction is $A_\mathrm{\textit{r'}} = 1.4\,\mathrm{mag}$ and $A_\mathrm{\textit{i'}} = 1.0\,\mathrm{mag}$; for \textit{z'} filter, we used the average extinction in \textit{i'} and \textit{J} bands, $A_\mathrm{\textit{z'}} = 0.74\,\mathrm{mag}$. Finally, the calibrated magnitudes were converted to flux densities using \citet{fukugita1996}. The complete dataset is available in Table \ref{tab:phot090727}.

\par The calibrated optical light curves are shown in Figure \ref{fig:lc}. The light curve has two main components: a sharp peak at early time with optical flux reaching a peak brightness of $0.17\,\mathrm{mJy}$ ($\mathrm{\textit{r'}} = 18.3\,\mathrm{mag}$) at around $t = 209\,\mathrm{s}$ and a broader, late time peak which reaches $\mathrm{\textit{r'}} \approx 20.2\,\mathrm{mag}$ before fading as a power law with decay index $\alpha \approx 0.7$. Below, we model the optical and X-ray light curves and test whether these components can be explained as reverse- and forward-shock components of the external shock (Type I afterglow from \citealt{gomboc2009}) or the early time emission originates from internal shocks that produce the gamma-ray peaks.

\section{GRB~090727 analysis and interpretation}
\label{sect:dataanal}

\begin{deluxetable*}{cccccc}[!h]
\tablecaption{GRB~090727: X-ray and optical light curves fit.}
\tabletypesize{\footnotesize}
\tablewidth{\linewidth}
\tablehead{
\colhead{Data set} &
\multicolumn{4}{c}{Fit parameters} & 
\colhead{$\chi _\mathrm{red} ^2$ (d.o.f.)} \\
\colhead{} &
\colhead{$\alpha _\mathrm{rise}$} & 
\colhead{$\alpha _\mathrm{decay}$} & 
\colhead{$t _\mathrm{break}$ [s]} & 
\colhead{$F_0$ [$10^{-4}\,\mathrm{Jy}$]} & 
\colhead{}
}
\startdata
X-ray peak & $-6.73 \pm 0.62$ & $16.37 \pm 2.83$ & $276.5 \pm 4.0$ & $3.0 \pm 0.2$ & $1.50$ (19)  \\
 & & & & & \\
 Optical (first peak) & $-7.81 \pm 2.19$ & $4.19 \pm 1.32$ & $216.9 \pm 15.2$ & $3.3 \pm 0.6$ & \\
 + & & & & & $1.01$ ($23$) \\
 Optical (second peak) & $-1.80 \pm 0.61$ & $0.70 \pm 0.06$ & $1191.5 \pm 162.2$ & $0.51 \pm 0.03$ & \\
 & & & & & \\
 Optical -- RS (first peak) & $-5.0\,^\ast$ & $2.0\,^\ast$ & $187.4 \pm 13.3$ & $1.5 \pm 0.2$ & \\
 + & & & & & $2.58$ ($27$) \\
 Optical -- FS (second peak) & $-0.5\,^\ast$ & $1.0\,^\ast$ & $3709.5 \pm 602.5$ & $0.43 \pm 0.05$ &
\enddata
\tablecomments{Early time X-ray data were fitted with the Beuermann profile (Eq. \ref{eq:beuermann}). Optical data were fitted with the sum of two contributions of the form of Eq. (\ref{eq:beuermann}). We also fitted the optical data assuming the reverse- (RS) and the forward- (FS) external shock scenario \citep{zhang2003} by fixing the power-law indices to the values suggested by the theory. Asterisks ($^\ast$) indicate fixed values.}
\label{tab:fit}
\end{deluxetable*}

\par We performed X-ray and optical LC modelling with a smoothly joint broken power-law fit \citep{beuermann1999}:
\begin{equation}
\label{eq:beuermann}
F(t) = F_0 \left[ \left( \frac{t}{t_\mathrm{break}} \right)^{\alpha _\mathrm{rise}} + \left( \frac{t}{t_\mathrm{break}} \right)^{\alpha _\mathrm{decay}} \right]^{-1}\,,
\end{equation}
from where
\begin{equation}
\label{eq:beuermann2}
t_\mathrm{peak} = t_\mathrm{break} \left( - \frac{\alpha_\mathrm{rise}}{\alpha_\mathrm{decay}} \right)^{1/(\alpha_\mathrm{decay}-\alpha_\mathrm{rise})}\,.
\end{equation}
The parameters are power-law rise and decay indices ($\alpha _\mathrm{rise}$ and $\alpha _\mathrm{decay}$), break time ($t_\mathrm{break}$) and flux density normalisation ($F_0$). We constrained the smoothness parameter ($n$) from the original equation to $n=1$, so that we have a simple broken power-law and fewer free parameters.

\par We fitted the X-ray peak using Eq. (\ref{eq:beuermann}) and the optical LC using the sum of two contributions of the form of Eq. (\ref{eq:beuermann}). By assuming a common origin for optical emission at different wavelengths, we could fit the optical data from all three optical bands simultaneously with common parameters, except normalisation. The results are presented in Table \ref{tab:fit}, where the flux density normalisation value ($F_0$) for the optical LC corresponds to $r'$ band data (shift factors obtained from the fit are $1.77 \pm 0.14$ for $i'$ band data and $2.69 \pm 0.37$ for $z'$ band data). The resulting functions are plotted in Figure \ref{fig:lc}. The X-ray emission at later times ($t > 4000 \,\mathrm{s}$) decays with a power-law index $\alpha_\mathrm{decay} = 0.79 \pm 0.05$ ($\chi_\mathrm{red}  ^2 = 0.72$, $6$ d.o.f.), if one excludes X-ray flares and assumes no jet break. This is consistent with the decay of the late time optical emission.

\subsection{Early time temporal properties}
\label{ssect:temporal_prop}

\par The peak times for the early time X-ray and optical LCs appear to be different in this modelling. From parameters in Table \ref{tab:fit} and from Eq. (\ref{eq:beuermann2}) we get $t_\mathrm{peak}^\mathrm{X-ray} = (266.1 \pm 4.1)\,\mathrm{s}$ and $t_\mathrm{peak}^\mathrm{opt} = (228.4 \pm 17.8)\,\mathrm{s}$. If we fix the optical break time and rise and decay indices to the ones obtained from the X-ray fit and force the optical peak time to be $266.1\,\mathrm{s}$, to check if the early time peaks are simultaneous, we do not improve the fit, but rather obtain the value of $\chi ^2 = 51.1$ for $28$ d.o.f., which gives the p-value of $0.0049$. If we allow the $1\sigma$ interval for the optical break time and rise and decay indices as obtained from the X-ray fit, we obtain $t_\mathrm{peak}^\mathrm{opt} = (263.2 \pm 0.7)\,\mathrm{s}$ and the value of $\chi ^2 = 34.7$ for $25$ d.o.f., which gives the p-value of $0.0934$. The p-value obtained when fitting the early time optical LC without fixing any of the parameters is $p = 0.4503$. 

\par These results suggest that it is most likely that early time X-ray and optical peaks are non-simultaneous; similarly, the decay power-law index of X-ray and optical peaks are inconsistent (Table \ref{tab:fit}), implying an uncommon origin. However, these discrepancies can be due to the sparse sampling of the optical light curve around the time of the X-ray peak or due to the model that we used to fit X-ray and optical peaks (Eq. \ref{eq:beuermann}). We therefore cannot exclude the possibility that the optical emission is simultaneous with the early time X-ray/gamma-ray peak, although we can reject the hypothesis that peaks are simultaneous at $90\%$ confidence level, based on the p-value.

\par If we assume that the early time optical peak is not from the same emission region as the gamma-ray and X-ray peaks at that time, we can test if the optical peak is perhaps due to external shock afterglow emission. We see that the optical LC in the $r$ band looks like a combination of the reverse-external (RS) and the forward-external shock (FS) emission, as described by \citet{zhang2003}. If the first peak is the contribution from the RS \citep{kobayashi2003}, the power-law indices should be $\alpha _\mathrm{rise} ^\mathrm{RS} = - 5$ and $\alpha _\mathrm{decay}^\mathrm{RS} = 2$. The second peak in this scenario is produced by the FS and the theory predicts that the power-law indices should be $\alpha _\mathrm{rise}^\mathrm{FS} = - 0.5$ and $\alpha _\mathrm{decay}^\mathrm{FS} = 1$ \citep{sari1998}. Following that, we fitted our optical data by fixing the values of indices to the theoretical values. Fit results are presented in Table \ref{tab:fit}, and it is evident that the overall LC shape is not fitted well with the RS/FS model (reduced $\chi ^2$ value from the RS/FS model is higher than for the case where no specific indices were assumed). Due to the steep decay index ($\alpha _\mathrm{decay} = 4.19$), which is steeper than the index for the high latitude emission ($\alpha = 2 + \beta$; this is the maximal value allowed in the external shock model) we rule out a simple RS origin of the early time optical emission.

\subsection{Early time spectral energy evolution}

\par We constructed the spectral energy distributions (SEDs) at three different epochs after the trigger (Figure \ref{fig:sed}). Three epochs that we used are indicated in Figure \ref{fig:lc} as vertical dashed lines. For the X-ray emission, we collected the data in time intervals indicated in Table \ref{tab:batxrt}: Epoch 1 is in time interval $t = [101 - 257]\,\mathrm{s}$ (XRT PC mode), Epoch 2 is in time interval $t = [260 - 300]\,\mathrm{s}$ (XRT WT mode) and Epoch 3 is in time interval $t = [4 \times 10^3 - 2 \times 10^6] \,\mathrm{s}$ (XRT PC mode). Then we renormalised the obtained flux density at the averaged time towards the time of interest (indicated in Table \ref{tab:xrtoptjoint}). Using \texttt{XSPECv12.6} software, we fitted the SMC-like extinction profile \citep{pei1992} to the joint X-ray and optical SEDs. We used the zdust$\ast$wabs$\ast$zwabs extinction model, where we fixed the extinction ratio to $\mathrm{R_V} = 2.93$. The best fits in Epochs 1 and 2 were obtained when we used the power-law model, and in Epoch 3 when we used the broken power-law model (for the broken power-law model we assumed $\beta_\mathrm{X} = \beta_\mathrm{opt} + 0.5$). The results of the fits are presented in Table \ref{tab:xrtoptjoint}. Because the redshift of GRB~090727 is not known, we fixed it to $z=0$ and the values of E(B-V) therefore represent the upper limits. When fitting all epochs simultaneously using the power-law model for Epochs 1, 2, and broken power-law model for Epoch 3, with E(B-V) as a common parameter, we obtain $\mathrm{E(B-V)} < 1.14$ ($\chi _\mathrm{red} ^2$ is $0.8$ with $61$ d.o.f.).

\begin{figure}[!h]
\includegraphics[width=1\linewidth]{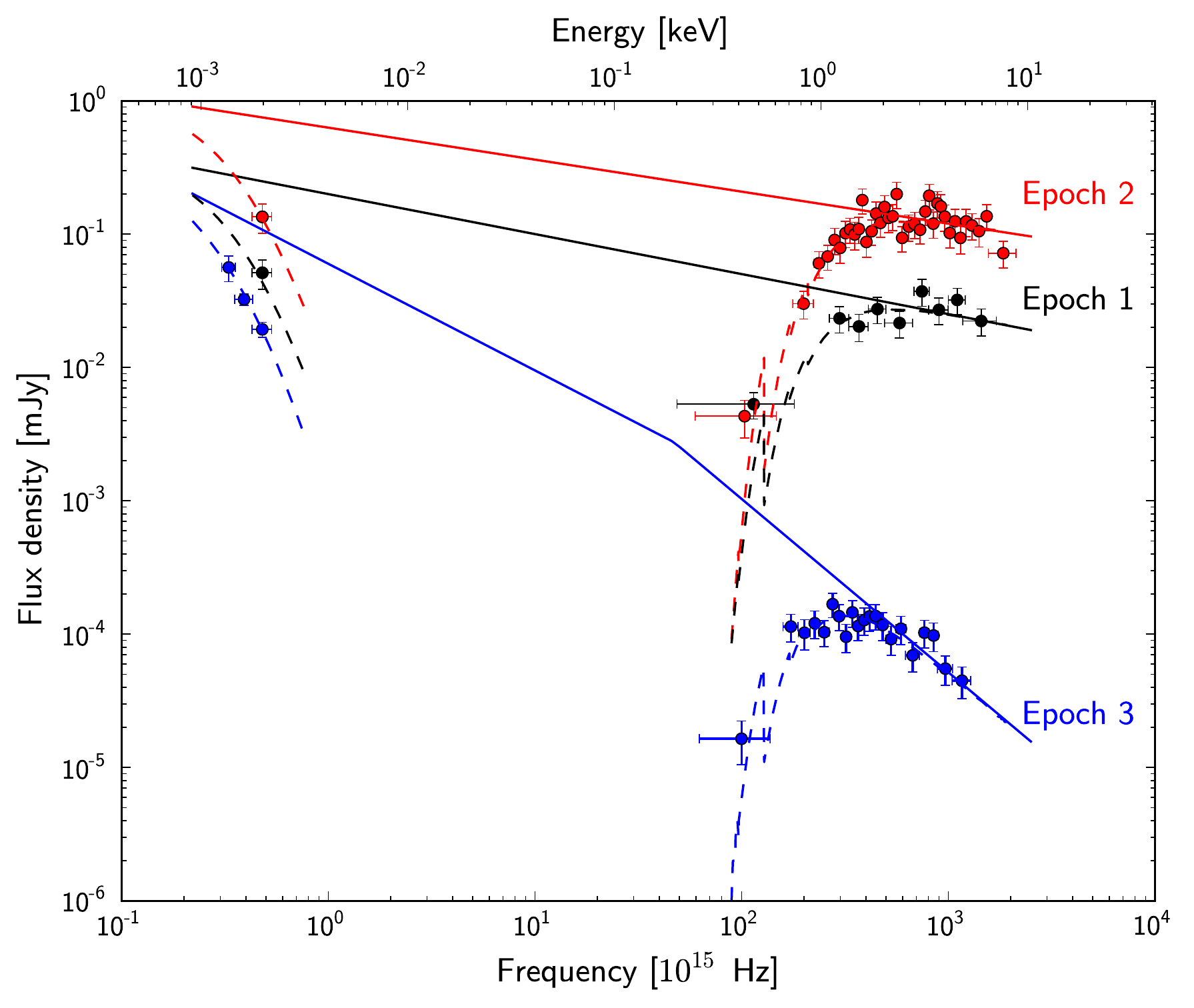} 
\caption{\label{fig:sed} Spectral energy distributions for three different epochs indicated in Figure \ref{fig:lc}. Joint X-ray and optical fits are plotted with solid lines.}
\end{figure}

\begin{deluxetable}{ccccc}[!htbp]
\tablecaption{GRB~090727: Joint X-ray and optical SEDs fit.}
\tabletypesize{\footnotesize}
\tablewidth{\linewidth}
\tablehead{
\colhead{Epoch} &
\colhead{Time} & 
\colhead{Spectral index} & 
\colhead{E(B-V)} &
\colhead{$\chi _\mathrm{red} ^2$ (d.o.f.)} \\
\colhead{} &
\colhead{$t \, [\mathrm{s}]$} & 
\colhead{$\beta$} & 
\colhead{} & 
\colhead{}
}
\startdata
1 (PL) & $160$ & $0.29_{-0.23}^{+0.24}$ & $< 1.55$ & $0.9$ ($7$) \\
2 (PL) & $267$ & $0.37_{-0.11}^{+0.11}$ & $< 1.69$ & $0.8$ ($35$) \\
3 (BPL) & $5150$ & $0.81_{-0.24}^{+0.28}$ & $< 1.10$ & $1.0$ ($25$) \\
  & & $1.31_{-0.24}^{+0.28}$ & &
\enddata
\tablecomments{Times since the GRB trigger are given for the optical data points. The time intervals used for fitting correspond to the ones from Table \ref{tab:batxrt}. In Epochs 1 and 2 the best results were obtained using the power-law (PL) model, while in Epoch 3 the broken power-law (BPL) model was used, combined with the extinction model (zdust$\ast$wabs$\ast$zwabs) and assuming SMC-like extinction. Values for E(B-V) are upper limits (at $z=0$). The break energy for the BPL fit for Epoch 3 is $E_\mathrm{break} = 0.19\,\mathrm{keV}$. The confidence intervals are at $90\%$.}
\label{tab:xrtoptjoint}
\end{deluxetable}

\par The spectral evolution is evident from SEDs, and the best fit models indicate the transition of the spectral break frequency. The X-ray to optical spectral index $\beta$ ($\beta = \Gamma - 1$, where $\Gamma$ is photon index) is smaller before and during the X-ray peak at $t \approx 267\,\mathrm{s}$ as compared to the value after the peak (Table \ref{tab:xrtoptjoint}). These values are consistent with the values reported in Table \ref{tab:batxrt} for the X-ray spectra.

\subsection{The origin of the optical emission}

\par In summary, we rule out the reverse-external shock origin for the early time optical emission in GRB~090727 and instead conclude that the early time optical flash likely corresponds to emission from an internal shock region. Although our modelling favours the non-simultaneous X-ray and optical peaks, due to the lack of observational data points around the peak time it is inconclusive. 

\par If the early time X-ray and optical peaks are simultaneous, they are caused by the same internal shock (the same collision inside the flow). On the other hand, if they are non-simultaneous, the peaks would originate from a slightly different locations in the flow (different internal shock components), possibly at different lab times and/or a different radii of the expanding flow.

\par We now examine other GRBs with contemporaneous gamma-ray and optical observations to test the origin of their emission components and put GRB~090727 in a broader observational and theoretical context.

\section{GRBs with early time optical emission}
\label{sect:sample}
\par Up to January 2012, there were at least $36$ GRBs with optical emission detected during prompt gamma-ray emission (Table \ref{tab:sampleallgrbs})\footnote{\label{fn_ref}References for data in Table \ref{tab:sampleallgrbs}: $\mathrm{T}_{90}$ was obtained from \citet{sakamoto2011}, except for GRB~990123 \citep{kippen1999}, GRB~041219A and GRB~080603A \citep{vianello2009}, GRB~060418 \citep{cummings2006}, GRB~060605 \citep{sato2006}, GRB~080319B \citep{racusin2008}, GRB~100901A \citep{sakamoto2010}, GRB~100906A \citep{barthelmy2010} and GRB~110205A \citep{markwardt2011}. For GRB~050820A the total duration is given instead of $\mathrm{T}_{90}$ \citep{cenko2006}. \\ Redshift and $E_\mathrm{iso}^{\gamma}$ references: [1] \citet{cusumano2006}, [2] \citet{depasquale2006}, [3] \citet{depasquale2007}, [4] \citet{ghirlanda2008}, [5] \citet{golenetskii2005}, [6] \citet{butler2006}, [7] \citet{stratta2009}, [8] \citet{ghirlanda2012}, [9] \citet{ruiz2007}, [10] \citet{perley2008}, [11] \citet{perley2008b}, [12] \citet{guidorzi2011}, [13] \citet{perley2011}. \\ Optical LC morphology references: given in the last column, except for GRB~050319 \citep{wozniak2005}, GRB~050401 \citep{rykoff2005}, GRB~050801 \citep{rykoff2006}, GRB~050904 \citep{boer2006}, GRB~051109A and GRB~051111 \citep{yost2007a}, GRB~060210 \citep{curran2007}, GRB~060418 and GRB~060607A \citep{molinari2007}, GRB~060605 \citep{ferrero2009}, GRB~061126 \citep{gomboc2008}, GRB~071010B \citep{wang2008} and GRB~081203A \citep{kuin2009}.}. This is a heterogeneous sample whose properties are determined by the start time of optical observations relative to the gamma-ray emission, the optical brightness, the density of sampling and the intrinsic light curve properties. 

\begin{deluxetable}{ccccccc}[!h]
\tablecaption{General properties of GRBs in the master sample, i.e., GRBs with optical detection during prompt gamma-ray emission.}
\tabletypesize{\footnotesize}
\tablewidth{\linewidth}
\tablehead{
\colhead{GRB} &
\colhead{$\mathrm{T}_{90} \, [\mathrm{s}]$} & 
\colhead{$z$} & 
\colhead{$E_\mathrm{iso,53}^{\gamma} \, [\mathrm{erg}]$} &
\colhead{LC} &
\colhead{Inc.} &
\colhead{Ref.}
}
\startdata
990123 & $63.3$ & $1.60$ & $16$ & a & yes & App. \ref{sect:a990123} \\
041219A & $460.0$ & $0.31^\ast$ & $1$ & a & yes & App. \ref{sect:a041219a} \\
050319 & $151.7$ & $3.24$ & $0.37$ & b & no & [1] \\
050401 & $33.3$ & $2.90$ & $3.5$ & b & no & [2] \\
050801 & $19.4$ & $1.56^\ast$ & $0.092$ & c & no & [3] \\
050820A	& $600.0$ & $2.61$ & $8.3$ & a & yes & App. \ref{sect:a050820a} \\
050904\footnote{The optical flare at $\sim 500\,\mathrm{s}$ after the trigger \citep{boer2006} occurs after the end of the prompt gamma-ray emission, but it is coincident with the X-ray flare, suggesting a common origin via the late internal shock model \citep{wei2006}.} & $181.7$ & $6.29$ & $12$ & c & no & [4] \\
051109A & $37.2$ & $2.35$ & $0.5$ & b & no & [5] \\
051111 & $64.0$ & $1.55$ & $0.62$ & b & no & [6] \\
060111B	& $58.8$ & $1.5 ^\ast$ & $0.32$ & b & no & [7] \\
060210 & $242.2$ & $3.91$ & $4.1$ & b & no & [8] \\
060418 & $52.0$ & $1.49$ & $1.3$ & c & no & [8] \\
060526 & $275.2$ & $3.22$ & $0.24$ & a & yes & App. \ref{sect:a060526}\\
060605 & $15.0$ & $3.77$ & $0.28$ & c & no & [8] \\
060607A & $103.0$ & $3.08$ & $1.1$ & c & no & [8] \\
060729 & $113.0$ & $0.54$ & $0.16$ & a & yes & App. \ref{sect:a060729} \\
060904B	& $171.9$ & $0.70$ & $0.02$ & a & yes & App. \ref{sect:a060904b} \\
060927 & $22.4$ & $5.47$ & $0.77$ & b & no & [9] \\
061007 & $75.7$ & $1.26$ & $10$ & a & yes & App. \ref{sect:a061007} \\
061121 & $81.2$ & $1.31$ & $2.8$ & a & yes & App. \ref{sect:a061121} \\
061126 & $50.3$ & $1.16$ & $1.1$ & b & no & [10] \\
071003\footnote{The optical bump at $\sim 250\,\mathrm{s}$ after the trigger is perhaps coincident with the late time gamma-ray activity (late internal shocks), but it is non-prominent when compared with the background afterglow component \citep{perley2008b}.} & $148.4$ & $1.60$ & $3.4$ & b & no & [11] \\
071010B & $36.1$ & $0.95$ & $0.21$ & c & no & [8] \\
080310 & $352.4$ & $2.43$ & $0.32$ & a & yes & App. \ref{sect:a080310} \\
080319B & $50.0$ & $0.94$ & $13$ & a & yes & App. \ref{sect:a080319b} \\
080603A\footnote{Early time optical peak is probably associated with the prompt emission \citep{guidorzi2011}, but this GRB was not included in the sub-sample due to sparsity of the data and too few optical data points during the peak to allow detailed temporal analysis.} & $150.0$ & $1.69$ & $0.22$ & a & no & [12] \\
080607 & $78.9$ & $3.04$ & $28$ & b & no & [13] \\
080810 & $107.7$ & $3.35$ & $3$ & a & yes & App. \ref{sect:a080810} \\
080905B & $101.6$ & $2.37$ & $0.24$ & a & yes & App. \ref{sect:a080905b} \\
080928 & $233.7$ & $1.69$ & $0.14$ & a & yes & App. \ref{sect:a080928} \\
081008 & $179.5$ & $1.97$ & $0.63$ & a & yes & App. \ref{sect:a081008} \\
081203A & $223.0$ & $2.05$ & $3.5$ & c & no & [8] \\
090727 & $302.0$ & / & / & a & yes & This paper \\
100901A & $439.0$ & $1.41$ & $0.63$ & a & yes & App. \ref{sect:a100901a} \\
100906A & $114.0$ & $1.73$ & $2.2$ & a & yes & App. \ref{sect:a100906a} \\
110205A & $257.0$ & $2.22$ & $4.6$ & a & yes & App. \ref{sect:a110205a}
\enddata
\tablecomments{\label{tab:sampleallgrbs}Columns are: GRB identifier, $\mathrm{T}_{90}$, redshift, $E_\mathrm{iso}^{\gamma}$ (in $10^{53}\,\mathrm{erg}$), LC morphology, included in the sub-sample (yes/no), references (given in footnote \ref{fn_ref}). Another event (GRB~080330) has an optical emission detected during gamma-ray emission \citep{guidorzi2009}, but it was not added here since it was classified as an X-ray flash. A strong optical flare also occurred at early times in GRB~080129, but it was not added to our sample because the optical emission was not detected during the gamma-ray emission \citep{greiner2009}. \\
Light curve morphology: (a) Optical peak/peaks during gamma-ray emission, (b) Optical decay during gamma-ray emission, (c) Optical rise during gamma-ray emission. \\
$^\ast$Redshift was not determined spectroscopically. The corresponding $E_\mathrm{iso}^{\gamma}$ was calculated at a given redshift.}
\end{deluxetable}

\par Based on the optical LC shape during gamma-ray emission we can define three categories: (a) Optical emission shows single or multiple peaks during gamma-ray emission, (b) Optical emission decays during gamma-ray emission and (c) Optical emission rises during gamma-ray emission and peaks after the end of gamma-ray emission. In Table \ref{tab:sampleallgrbs} we summarise the properties of the master sample including qualitative indications of light curve morphology. Figure \ref{fig:sample_z_eiso} shows redshift and $E_\mathrm{iso}^{\gamma}$ distributions for GRBs from Table \ref{tab:sampleallgrbs}, which have an optical detection during gamma-ray emission. We see that the distributions appear consistent with a general GRB population when comparing them with, for example, Figure 10 from \citet{gomboc2012} and Figure 4 from \citet{ghirlanda2012}. The distributions appear consistent also for a subgroup of GRBs which show optical peak/peaks during gamma-ray emission (category (a) optical LCs) and which are studied in this paper.

\begin{figure}[!h]
\includegraphics[width=1\linewidth]{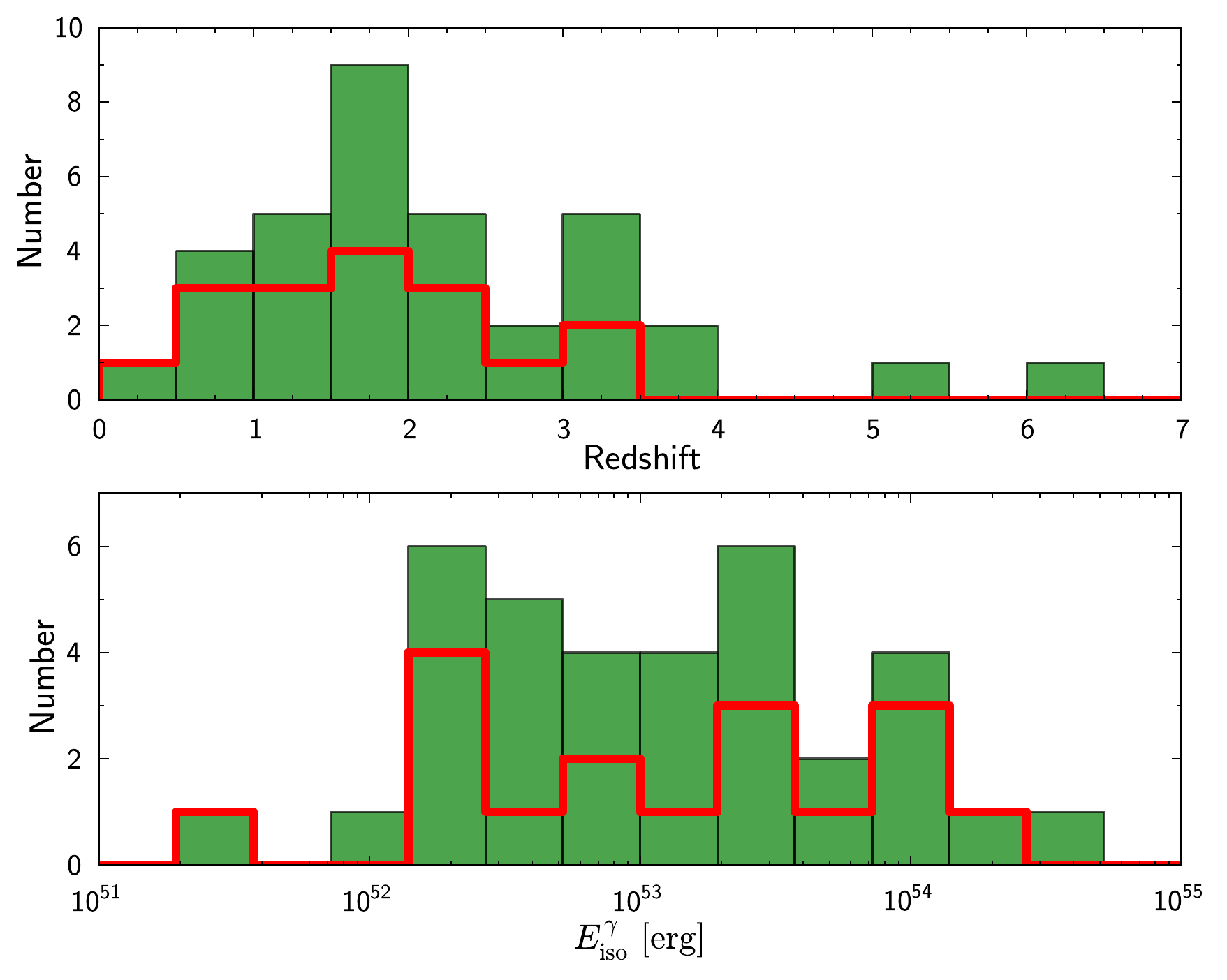} 
\caption{\label{fig:sample_z_eiso}Redshift and $E_\mathrm{iso}^{\gamma}$ distributions for $35$ GRBs which show optical detection during prompt gamma-ray emission (green bars). Red line represents $17$ GRBs for which optical LCs show peaks during gamma-ray emission and which were studied in the sub-sample. GRB~090727 is not included since no redshift and $E_\mathrm{iso}^{\gamma}$ are available.}
\end{figure}

\par In order to perform a detailed temporal analysis to investigate more thoroughly the origin of early time optical emission, we select a heterogeneous sub-sample of $18$ GRBs that show behaviour similar to GRB~090727, i.e., where early time optical emission displays single or multiple peaks during gamma-ray emission (category (a)). Various interpretations for GRBs with early time optical emission have been proposed in the literature, and by studying the sample of such GRBs we aim to investigate whether there are any similarities amongst them. Similar studies of GRBs with early time optical detections (and with early time optical upper limits) were presented in \citet{yost2007b} and \citet{yost2007a}, as well as in \citet{panaitescu2008}, but there they concentrated mostly on the spectral properties. In this paper we additionally explore the temporal domain, focusing attention on GRBs for which early time optical emission shows peaks during gamma-ray emission (Figures \ref{fig:samplelcs} and \ref{fig:lc_rf}).

\par In the Appendix \ref{sect:aa} (\ref{sect:a990123}-\ref{sect:a110205a}) we describe the following GRBs from the sub-sample that we study together with GRB~090727: GRB~990123, GRB~041219A, GRB~050820A, GRB~060526, GRB~060729, GRB~060904B, GRB~061007, GRB~061121, GRB~080310, GRB~080319B, GRB~080810, GRB~080905B, GRB~080928, GRB~081008, GRB~100901A, GRB~100906A and GRB~110205A. Information gathered from the literature can help us understand the nature of the early time optical emission for each case in more detail. In the main part of this paper we concentrate on epochs during gamma-ray emission and on temporal and spectral characteristics. However, in the Appendix \ref{sect:aa} we also summarise the properties of the late time afterglow behaviour, GRB energetics and optical light curve shapes and characteristics, to obtain a broader picture.

\begin{figure*}[!ht]
\includegraphics[width=1\linewidth]{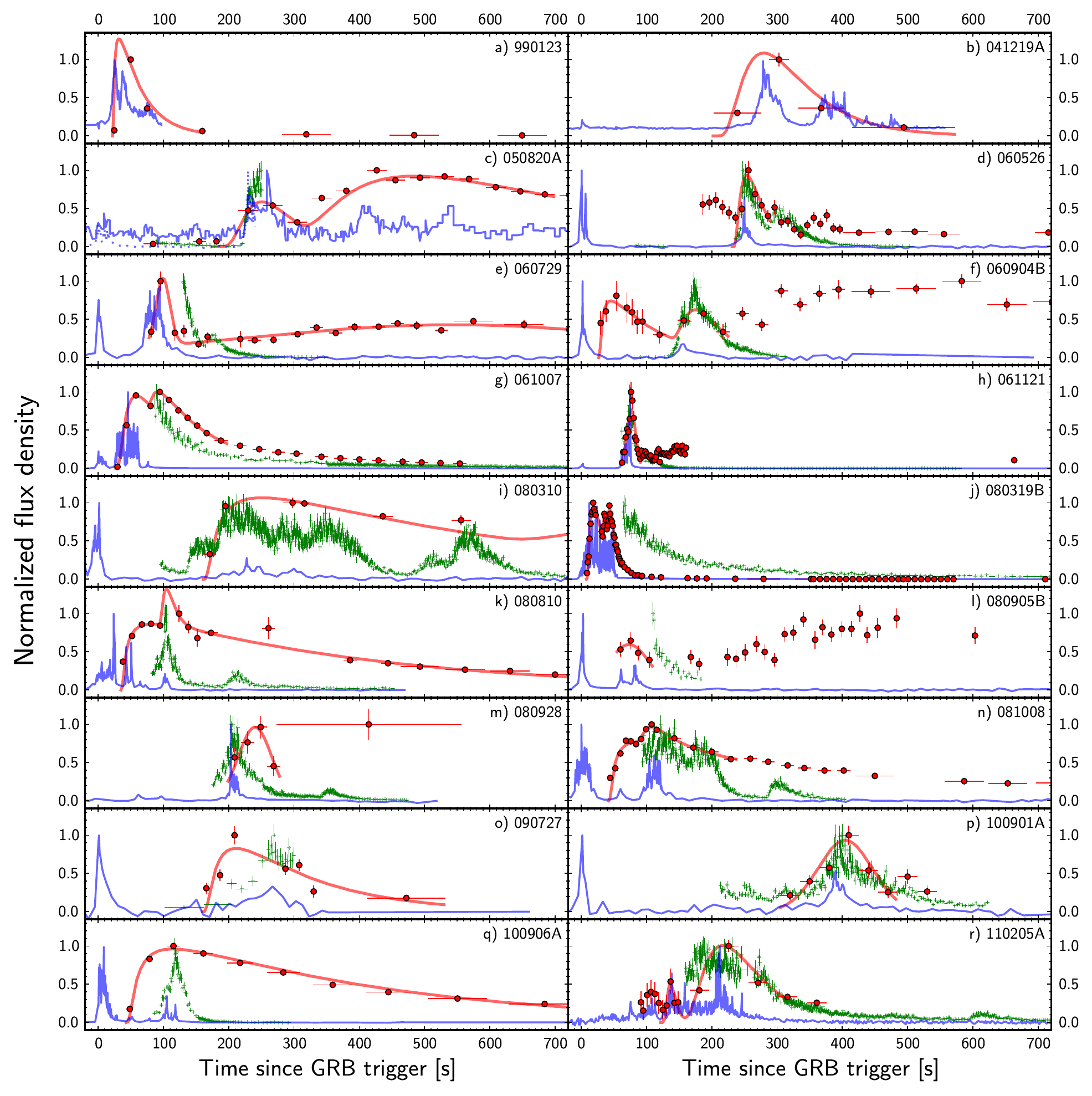} 
\caption{\label{fig:samplelcs} Gamma-ray (light blue solid line), X-ray (green crosses) and optical (red points) normalised light curves for GRBs from the sub-sample. Red solid line represents the best optical peak profile fit (Norris or Gaussian), with parameters taken from Table \ref{tab:sampletaboptical}. \\
Optical emission references and filter bands are given in Appendices \ref{sect:a990123}-\ref{sect:a110205a}. X-ray emission references: all from the XRT lightcurve repository \citep{evans2009b}. Gamma-ray emission references: a) BATSE archive, $100-300\,\mathrm{keV}$; b) BAT archive, $15-100\,\mathrm{keV}$; c) solid line: KONUS, $18-70\,\mathrm{keV}$; dotted line: BAT archive, $15-150\,\mathrm{keV}$; f) BAT archive, $10-20\,\mathrm{keV}$; c), d), e), g), h), i), j), k), l), m), n), o), p), q) and r) BAT archive, $15-150\,\mathrm{keV}$. Gamma-ray energy bands and time bins are chosen in a way to best represent the LCs features.}
\end{figure*}
\begin{figure*}[!ht]
\includegraphics[width=0.5\linewidth]{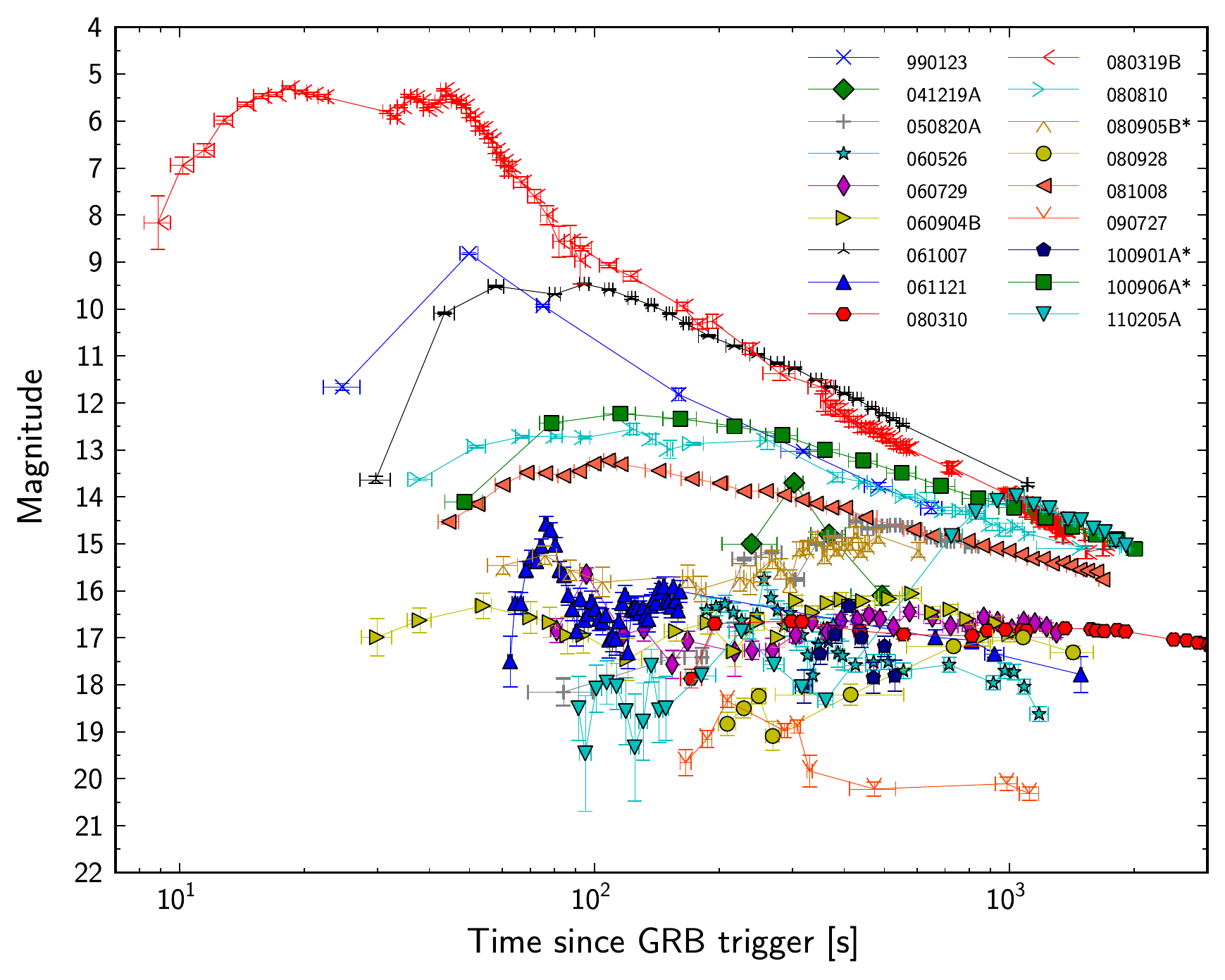}
\includegraphics[width=0.5\linewidth]{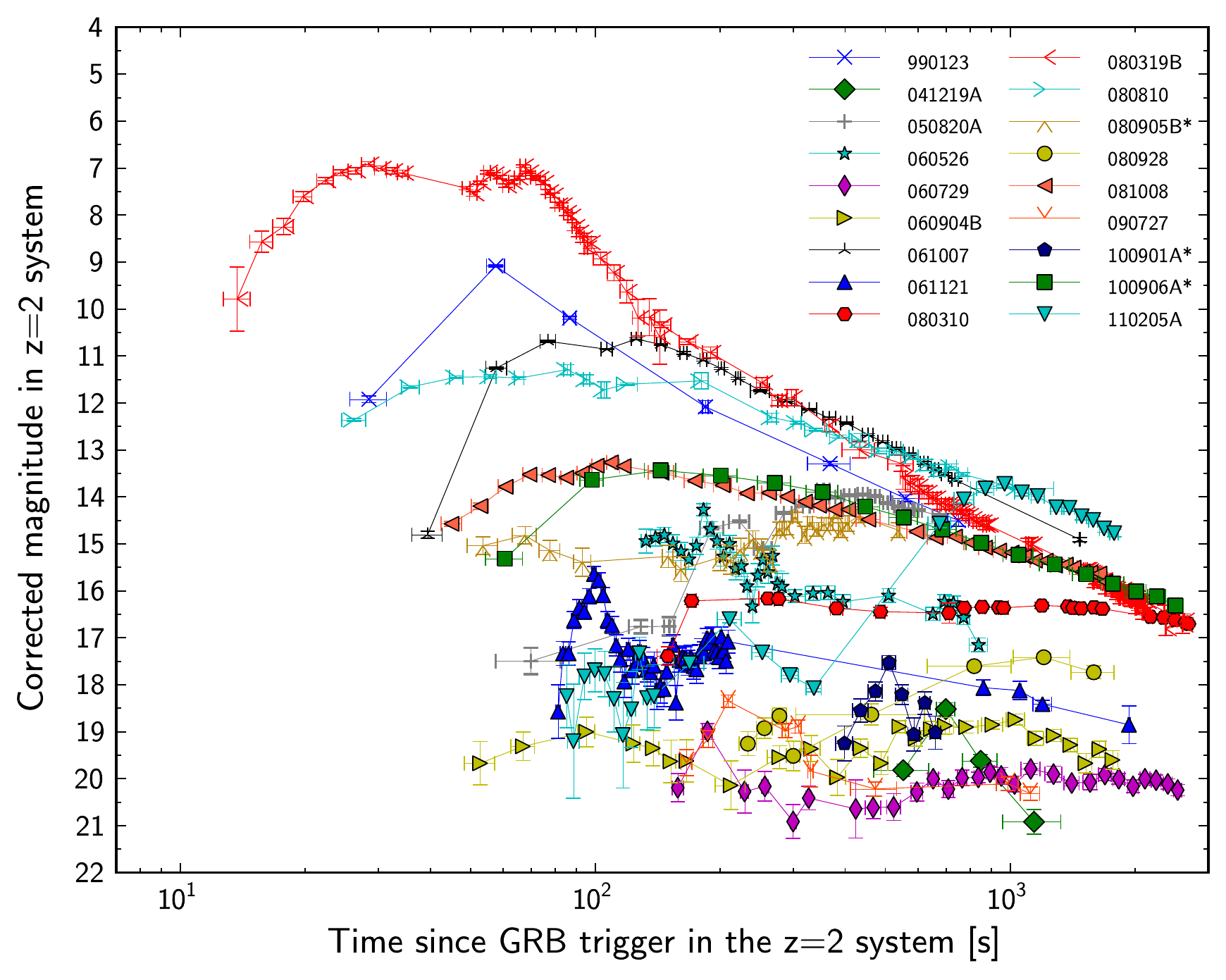}
\caption{\label{fig:lc_rf} Early time optical light curves in the observer (left) and in the $z=2$ (right) reference frame for GRBs in the sub-sample. Magnitudes are corrected for the Galactic extinction, but not for the host galaxy extinction. For GRB~090727 we could not scale the light curve due to unknown redshift, while for GRB~041219A and GRB~060111B we used redshifts provided in Table \ref{tab:sampleallgrbs}. For GRB~080905B and GRB~100901A the magnitudes are preliminary (\ref{sect:a080905b}, \ref{sect:a100901a}).}
\end{figure*}

\section{Sample discussion}
\label{sect:sampledisc}
\par The sample of GRBs which show optical peaks during their gamma-ray emission is clearly inhomogeneous, as inferred from the shape of the light curves at early and late times (see Figures \ref{fig:samplelcs} and \ref{fig:lc_rf}), as well as from the various different theoretical interpretations presented in the Appendix \ref{sect:aa}. Our motivation is to try to examine in detail all the properties that we can obtain during early phases of GRBs and determine whether there exists a common model to interpret prompt optical emission. 

\subsection{Temporal properties}
\par Various mechanisms can produce features in the temporal domain of early time optical emission. In many cases, optical LCs at early times are very sparsely sampled, causing difficulty in interpreting the behaviour. It is possible to simply discuss the morphology of the LCs: several LCs show one or more optical peaks during gamma-ray emission (category (a) from Section \ref{sect:sample}), while many LCs show only the power-law rise or decay behaviour. This is clearly an observational bias, since optical observations can start too late and the peak onset is missed, or gamma-ray emission is relatively short-lived and the peak occurs after the end of prompt gamma-ray emission. When dealing with LCs that show optical peaks during gamma-ray emission, it is sometimes non-trivial to determine the peak time and the rise and decay indices, especially when there are only a few data points. Nevertheless, it is interesting to investigate whether gamma-ray and optical emission are simultaneous, which would require that they peak at the same time and have similar rise and decay slopes. This would indicate that we may be dealing with radiation across the spectrum coming from a single emission site. 

\par For a number of GRBs it seems that optical LC tracks gamma-ray and/or X-ray emission (e.g., GRB~060526 -- App.~\ref{sect:a060526}; GRB~061121 -- App.~\ref{sect:a061121}; GRB~080319B -- App.~\ref{sect:a080319b}; GRB~110205A -- App.~\ref{sect:a110205a}), but it is not trivial to confirm the temporal correlation due to insufficient optical coverage or inconsistent power-law indices in most cases. On the other hand, there are several cases where early time optical emission is clearly not directly correlated with gamma-ray emission or where optical emission is too poorly sampled to see any clear correlation (e.g., GRB~990123 -- App.~\ref{sect:a990123}; GRB~080928 -- App.~\ref{sect:a080928}). \citet{rykoff2009} studied $12$ GRBs with early optical detection and concluded that for none of them show obvious correlation between optical and gamma-ray emission.

\par From the available literature (for the references see Appendices \ref{sect:a990123}-\ref{sect:a110205a}) we gathered early time photometric data for GRBs in the sub-sample. In Figure \ref{fig:samplelcs} we plot all of the LCs with normalised flux density in gamma-ray, X-ray, and optical bands. We also plot in Figure \ref{fig:lc_rf} all early time optical LCs, both in the observer's frame and in the rest frame at the average redshift of $z=2$ (using Eq. 1 from \citealt{panaitescuvestrand2008}). We can clearly see the diversity of LCs at early times. There is no clear ``canonical'' early time optical LC. At $\sim 1000\,\mathrm{s}$ after the trigger, optical LCs start to show the canonical afterglow shape, with comparable and consistent decay indices.

\par GRBs studied in this paper are long duration GRBs. It is clear that long duration is crucial to obtain optical measurements during gamma-ray emission, but this is an observation bias. We further noticed that many GRBs in the sub-sample have very similar gamma-ray emission LC: when initial gamma-ray peak decays, it is followed by the second and sometimes even the third peak. In some cases there is extended gamma-ray emission present after the initial peak. This behaviour lasts for tens or sometimes even hundreds of seconds. It is thus not trivial to interpret early time optical emission and its origin, especially while it may be due to the standard afterglow emission produced by the initial gamma-ray peak, especially when optical LC shows smooth and non-steep behaviour.

\par To analyse the temporal properties of optical peaks, namely the duration ($w$) and the time when a peak reaches its maximum ($t_\mathrm{peak}$), we first fitted the Norris profile \citep{norris2005}, which is usually applied when dealing with peaks that show fast rise and exponential decay, although it is not necessary a true physical model. The profile takes the form \citep{norris2005}:
\begin{equation}
\label{eq:norris}
I(t) = A \, \exp{\left(2\sqrt{\frac{\tau _1}{\tau _2}}\right)}\,\exp{\left(-\frac{\tau _1}{t} - \frac{t}{\tau_2}\right)}\,,
\end{equation}
where $A$ is the normalisation constant and $\tau _1$ and $\tau _2$ are factors that control the shape of the pulse, however they are not directly rise and decay constant, but their combination across the whole pulse's duration influences the overall profile. From these parameters one can determine the peak time $t_\mathrm{peak} = \sqrt{\tau_1 \tau_2}$ and the peak duration $w = \tau _2 \, \sqrt{1 + 4\sqrt{\tau _1 /\tau _2}}$ \citep{norris2005}. Besides the Norris profile, we also fitted the Gaussian profile (following \citealt{burrows2006} and \citealt{kruhler2009}) in the form:
\begin{equation}
I(t) = A \, \exp{\left(\frac{-(t-t_\mathrm{peak})^2}{2\sigma^2}\right)}\,,
\end{equation}
where $\sigma = \mathrm{FWHM}/(2\sqrt{2\ln 2})$ and we estimated the duration of the peak as $w = 2 \times \mathrm{FWHM}$. Although this model is even less physically motivated than the Norris model, it provides a reasonable fit in a few cases. Following \citet{chincarini2010}, we calculated the typical relative time-scales of variations for optical peaks ($w/t_\mathrm{peak}$ or $\Delta t/t$). The results are presented in Table \ref{tab:sampletaboptical}. 

\begin{deluxetable*}{cccccccccc}[!h]
\tablecaption{Early time optical peaks temporal properties.}
\tabletypesize{\footnotesize}
\tablewidth{\linewidth}
\tablehead{
\colhead{} & 
\multicolumn{3}{c}{Norris fit} & 
\multicolumn{3}{c}{Gaussian fit} & 
\colhead{} &
\multicolumn{2}{c}{Variability} 
\\
\colhead{GRB} & 
\colhead{Peak time} & 
\colhead{Duration} &  
\colhead{} &  
\colhead{Peak time} & 
\colhead{Duration} &  
\colhead{} & 
\colhead{Interval} & 
\colhead{Norris} &
\colhead{Gaussian}
\\
\colhead{} & 
\colhead{$t_\mathrm{peak}^\mathrm{Norris}$ [s]} & 
\colhead{$w^\mathrm{Norris}$ [s]} &  
\colhead{$\chi ^2 _\mathrm{red}$} & 
\colhead{$t_\mathrm{peak}^\mathrm{Gauss}$ [s]} & 
\colhead{$w^\mathrm{Gauss}$ [s]} &  
\colhead{$\chi ^2 _\mathrm{red}$} & 
\colhead{[s]} & 
\colhead{$\frac{w^\mathrm{Norris}}{t_\mathrm{peak}^\mathrm{Norris}}$} &
\colhead{$\frac{w^\mathrm{Gauss}}{t_\mathrm{peak}^\mathrm{Gauss}}$} 
} 
\startdata
990123 & $31.7 \pm 2.4$ & $51.6 \pm 4.4$ & $19.1$ & / & / & / & $22$-$162$ & $1.63 \pm 0.19$ & / \\
041219A & $279.6 \pm 50.8$ & $147.6 \pm 72.4$ & $0.5$ & / & / & / & $202$-$573$ & $0.53 \pm 0.28$ & / \\
050820A & $250.8 \pm 33.5$ & $109.2 \pm 62.1$ & $6.7$ & / & / & / & $176$-$700$ & $0.44 \pm 0.25$ & / \\
& $488.2 \pm 19.3$ & $594.3 \pm 65.2$ & & / & / &  & $290$ & $1.22 \pm 0.14^\dag$ & / \\
060526 & $253.7 \pm 4.2$ & $50.1 \pm 14.3$ & $1.1$ & / & / & / & $231$-$291$ & $0.20 \pm 0.06$ & / \\
060729 & $88.9 \pm 3.5$ & $58.0 \pm 11.9$ & $1.5$ & $99.2 \pm 3.1$ & $49.8 \pm 11.1$ & $1.1$ & $78$-$752$ & $0.65 \pm 0.14$ & $0.50 \pm 0.11$ \\
& $486.6 \pm 68.2$ & $1320.3 \pm 421.8$ & & $544.9 \pm 21.8$ & $1482.9 \pm 164.8$ & & $151$ & $2.71 \pm 0.95^\dag$ & $2.72 \pm 0.32^\dag$ \\
060904B & $45.4 \pm 10.0$ & $103.4 \pm 55.7$ & $0.3$ & / & / & / & $27$-$227$ & $2.28 \pm 1.32$ & / \\
& $177.5 \pm 19.8$ & $78.5 \pm 58.7$ & & / & / & & $127$ & $0.44 \pm 0.33$ & / \\
061007 & $59.1 \pm 3.6$ & $99.5 \pm 20.5$ & $1.0$ & / & / & / & $27$-$199$ & $1.68 \pm 0.36$ & / \\
& $108.1 \pm 7.7$ & $173.3 \pm 46.9$ & & / & / & & $77$ & $1.60 \pm 0.45^\dag$ & / \\
061121 & $73.1 \pm 1.2$ & $21.4 \pm 2.7$ & $3.4$ & $76.7 \pm 0.4$ & $31.8 \pm 1.9$ & $1.0$ & $61$-$92$ & $0.29 \pm 0.04$ & $0.41 \pm 0.02$ \\
080310 & $250.8 \pm 13.3$ & $583.5 \pm 53.7$ & $0.7$ & / & / & / & $161$-$3100$ & $2.33 \pm 0.25$ & / \\
 & $1589 \pm 93$ & $4756 \pm 493$ & & / & / & & $571$ & $2.99 \pm 0.36^\dag$ & / \\
080319B & $18.8 \pm 0.9$ & $32.1 \pm 3.2$ & $1.4$ & / & / & / & $8$-$62$ & $1.65 \pm 0.18$ & / \\
& $36.3 \pm 1.0$ & $6.6 \pm 2.1$ & & / & / & & $32$ & $0.18 \pm 0.06$ & / \\
& $44.5 \pm 0.6$ & $14.0 \pm 1.6$ & & / & / & & $38$ & $0.31 \pm 0.04$ & / \\
080810 & $83.0 \pm 4.3$ & $451.3 \pm 20.4$ & $0.9$ & / & / & / & $35$-$730$ & $5.44 \pm 0.37^\dag$ & / \\
& $104.8 \pm 8.0$ & $24.7 \pm 14.1$ & & / & / &  & $92$ & $0.24 \pm 0.14$ & / \\
080905B & $67.5 \pm 9.7$ & $67.3 \pm 50.4$ & $0.2$ & $73.4 \pm 8.9$ & $147.4 \pm 79.5$ & $0.6$ & $55$-$110$ & $1.00 \pm 0.76$ & $2.01 \pm 1.11$ \\
080928 & $233.3 \pm 14.7$ & $78.0 \pm 43.2$ & $2.3$ & $240.3 \pm 6.1$ & $119.8 \pm 40.7$ & $0.6$ & $199$-$279$ & $0.33 \pm 0.19$ & $0.50 \pm 0.17$ \\
081008 & $89.7 \pm 4.7$ & $399.7 \pm 53.9$ & $1.4$ & / & / & / & $42$-$240$ & $4.46 \pm 0.64^\dag$ & / \\
& $109.4 \pm 10.9$ & $58.0 \pm 35.3$ & & / & / & & $81$ & $0.53 \pm 0.33$ & / \\
090727 & $211.4 \pm 8.0$ & $202.6 \pm 44.5$ & $3.6$ & / & / & / & $160$-$531$ & $0.96 \pm 0.21$ & / \\
100901A & $397.1 \pm 19.9$ & $98.1 \pm 32.1$ & $4.1$ & $404.2 \pm 6.5$ & $195.9 \pm 28.1$ & $0.8$ & $305$-$485$ & $0.25 \pm 0.08$ & $0.48 \pm 0.07$ \\
100906A & $114.1 \pm 6.7$ & $464.9 \pm 28.7$ & $1.2$ & / & / & / & $43$-$911$ & $4.07 \pm 0.35^\dag$ & / \\
110205A & $136.7 \pm 10.3$ & $18.5 \pm 17.5$ & $1.0$ & / & / & / & $122$-$331$ & $0.14 \pm 0.13$ & / \\
& $217.1 \pm 27.8$ & $126.9 \pm 46.3$ & & / & / & & $140$ & $0.58 \pm 0.23$ & /  
\enddata
\tablecomments{Peak time, peak duration and relative variability time-scale for GRBs from the sub-sample (Figure \ref{fig:samplelcs}), as obtained from fitting the Norris and the Gaussian profile in the specified time interval. The Gaussian profile is only given if it improves the fit. The start time for the Norris profile fitting equals the start time of the interval for the first peak, while for the next peaks only the start time is given in the interval column. $^\dag$ indicates that the values probably do not represent optical flashes, but rather the domination of the afterglow light (forward-external shock; see the text and the references in App. \ref{sect:aa}).}
\label{tab:sampletaboptical}
\end{deluxetable*}

\par The results from fitting the Gaussian profile are only given if they provide a better fit. Furthermore, we note that the results of the fit are in some cases indicative of a bad model, but this could also be due to the heterogeneous sample (especially non-homogeneous photometric calibration of optical observations). Although the values for the duration and the peak time are reasonable, the derived uncertainties in such cases can be underestimated. Especially for GRBs where there are only a few data points and the uncertainties obtained from the photometry are very small (e.g., GRB~990123 and GRB~050820A), this leads to a very high $\chi^2$ of the fit. Due to consistency with our analysis we decided to keep the results despite the poor fit.

\par Figure \ref{fig:dtt} shows the number of optical peaks having certain ratio between the duration and the time of the peak, $\Delta t/t$ ($=w/t_\mathrm{peak}$). We used the best fit values for each GRB from Table \ref{tab:sampletaboptical}. We did not plot the values for the peaks which could be affected by the afterglow emission, as inferred from their smooth and non-steep light curve behaviour typically at later times or from the literature (App. \ref{sect:aa}). These values are marked with $^\dag$ in Table \ref{tab:sampletaboptical}. 

\begin{figure}[!h]
\includegraphics[width=1\linewidth]{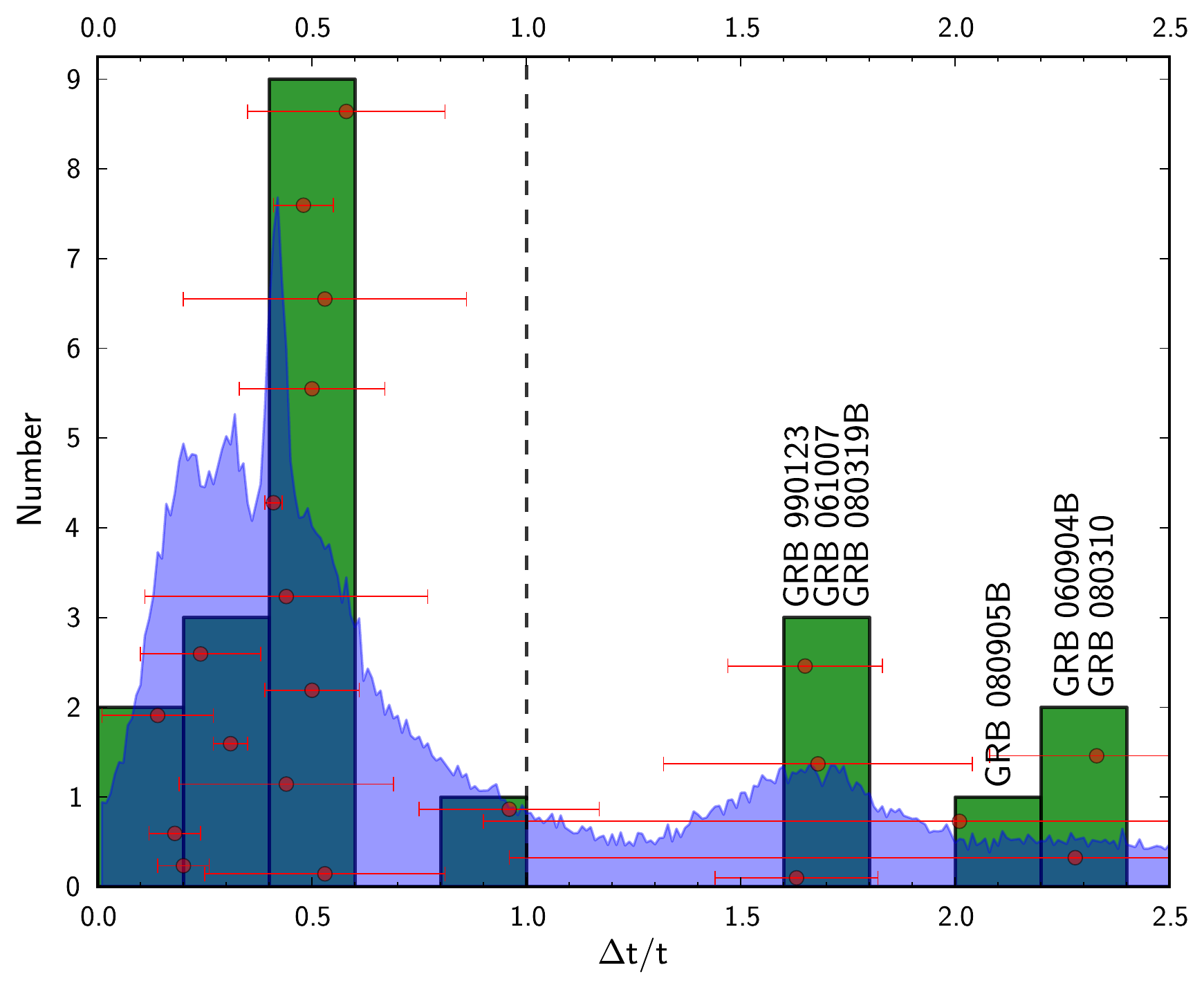}
\caption{\label{fig:dtt} Distribution of the ratio between duration and peak time ($w/t_\mathrm{peak}$ or $\Delta t/t$) for early time optical peaks. Data are obtained from Table \ref{tab:sampletaboptical} using the best fit values for each GRB (red points and the corresponding green histogram). Light blue normalised histogram in the background represents a Monte Carlo simulation, which takes into account also the $\Delta t/t$ error bars (associated with red points).}
\end{figure}

\par When plotting the histogram of the $\Delta t/t$ distribution, care must be taken as there are large $\Delta t/t$ error bars in several cases. To take into account these error bars in the histogram, we performed the following Monte Carlo simulation: for each GRB from the sub-sample we chose a $\Delta t/t$ value, as obtained from the data (Table \ref{tab:sampletaboptical}). We added/subtracted the corresponding $\Delta t/t$ deviation, which is also obtained from the data, but we multiply this deviation with the normally distributed random coefficient:
\begin{equation}
\label{eq:mc}
\left(\frac{\Delta t}{t}\right)^\mathrm{simulation} = \left(\frac{\Delta t}{t}\right)^\mathrm{data} + \epsilon \times \delta \left[ \left(\frac{\Delta t}{t}\right)^\mathrm{data} \right]\,,
\end{equation}
where $\epsilon$ represents a random number drawn from a normal distribution with zero mean and unit variance and $\delta[\Delta t/t]$ is the error, associated with the mean value of $\Delta t/t$ from Table \ref{tab:sampletaboptical}. We generated $10^5$ simulated $\Delta t/t$ values by this method and plotted the normalised histogram in Figure \ref{fig:dtt} (light blue color in the background). With this method we can check when the error bars should be considered with care. For example, we see in Figure \ref{fig:dtt} that GRB~990123, GRB~061007 and GRB~080319B make a significant bump at $\Delta t/t > 1$ (although $\Delta t/t$ for the peak of GRB~061007 might be affected by the afterglow emission and does not necessary represent only the early time optical flash). On the other hand, for GRB~080905B and GRB~060904B we can not confirm the high $\Delta t/t$ because of their large error bars.

\par In Table \ref{tab:plindices} and Figure \ref{fig:risedecay} we present the power-law rise and decay indices as reported in the literature or obtained from the data by fitting the simple power-law or the Beuermann profile (Eq. \ref{eq:beuermann}). References are given in Appendices \ref{sect:a990123}-\ref{sect:a110205a}. In Figure \ref{fig:risedecay} the radius of the circle around each data point represents the relative time-scale of the variations -- larger radius means higher $\Delta t/t$.

\begin{deluxetable}{cccc}[!h]
\tablecaption{Early time optical peaks power-law indices.}
\tablewidth{\linewidth}
\tabletypesize{\footnotesize}
\tablehead{
\colhead{GRB} &
\colhead{$\alpha _\mathrm{rise}$} & 
\colhead{$\alpha _\mathrm{decay}$} &
\colhead{$t _\mathrm{peak}$ [s]} 
}
\startdata
\textbf{990123} & $< -3.5$ & $2.1 \pm 0.2$ & $\sim 50$ \\
041219A & $\sim -5.4$ & $\sim 4.6$ & $303 \pm 15$ \\
\textbf{050820A} & / & $\approx 1.0\,^\dag$ & $\sim 425 \,\mathrm{s}$ \\
\textbf{060526} & $-12.8 \pm 3.0$ & $5.8 \pm 0.8$ & $256 \pm 5$ \\
060729 & $\sim -6.7$ & $\sim 3.7$ & $96 \pm 3$ \\
\textbf{060904B} & $< -0.2$ & $>0.8$ & $53 \pm 11$ \\
\textbf{061007} & $\sim -9$ & / & $\sim 58$ \\
\textbf{061007} & / & $1.7 \pm 0.1$ & $\sim 95$ \\
\textbf{061121} & $-12.6 \pm 2.0$ & $17.9 \pm 3.1$ & $77 \pm 1$ \\
\textbf{080310} & $-3.9 \pm 1.0$ & $2.8 \pm 0.9$ & $255 \pm 14$ \\
\textbf{080310} & $-1.0 \pm 0.5 \,^\dag$ & $1.0 \pm 0.4 \,^\dag$ & $1337 \pm 553$ \\
\textbf{080319B} & $-4.6 \pm 0.7$ & / & $18 \pm 1$ \\
\textbf{080319B} & / & $6.5 \pm 0.9$ & $43 \pm 1$ \\
\textbf{080810} & $-1.3 \pm 0.1 \,^\dag$ & $1.2 \pm 0.1\,^\dag$ & $\sim 100$ \\
080905B & $\sim -1.1$ & $\sim 2.3$ & $\sim 75$ \\
080928 & $\sim -3.0$ & $ \sim 9.8$ & $\sim 250$ \\
\textbf{081008} & $-5.7 \pm 1.6$ & $0.8 \pm 0.1$ & $76 \pm 6$ \\
\textbf{081008} & / & $0.8 \pm 0.2 \,^\dag$ & $111 \pm 5$ \\
\textbf{090727} & $-7.8 \pm 2.2$ & $4.2 \pm 1.3$ & $228 \pm 18$ \\
\textbf{100901A} & $-11.2 \pm 4.4$ & $7.9 \pm 2.6$ & $406 \pm 23$ \\
\textbf{100906A} & $-2.9 \pm 0.5$ & $1.0 \pm 0.1$ & $124 \pm 9$ \\
\textbf{110205A} & $-7.6 \pm 3.4$ & $3.4 \pm 0.9$ & $212 \pm 24$
\enddata
\tablecomments{\label{tab:plindices}Rise and decay power-law indices for GRBs in the sub-sample, as obtained by assuming the peak time from the last column. References are given in Appendices \ref{sect:a990123}-\ref{sect:a110205a}. For GRB~041219A, GRB~060729, GRB~080905B and GRB~080928 there are too few data to perform the fit, so the power-law indices are rough estimates (estimated from only two data points, as explained in Appendix \ref{sect:aa} for each GRB). For GRB~061007, GRB~080319B and GRB~081008 two peak times are given because the LC is composed of at least two well separated optical peaks. $^\dag$ indicates that the power-law index is associated with the afterglow and not directly with the optical flash; these values are not presented in Figure \ref{fig:risedecay}.}
\end{deluxetable}

\begin{figure}[!h]
\includegraphics[width=1\linewidth]{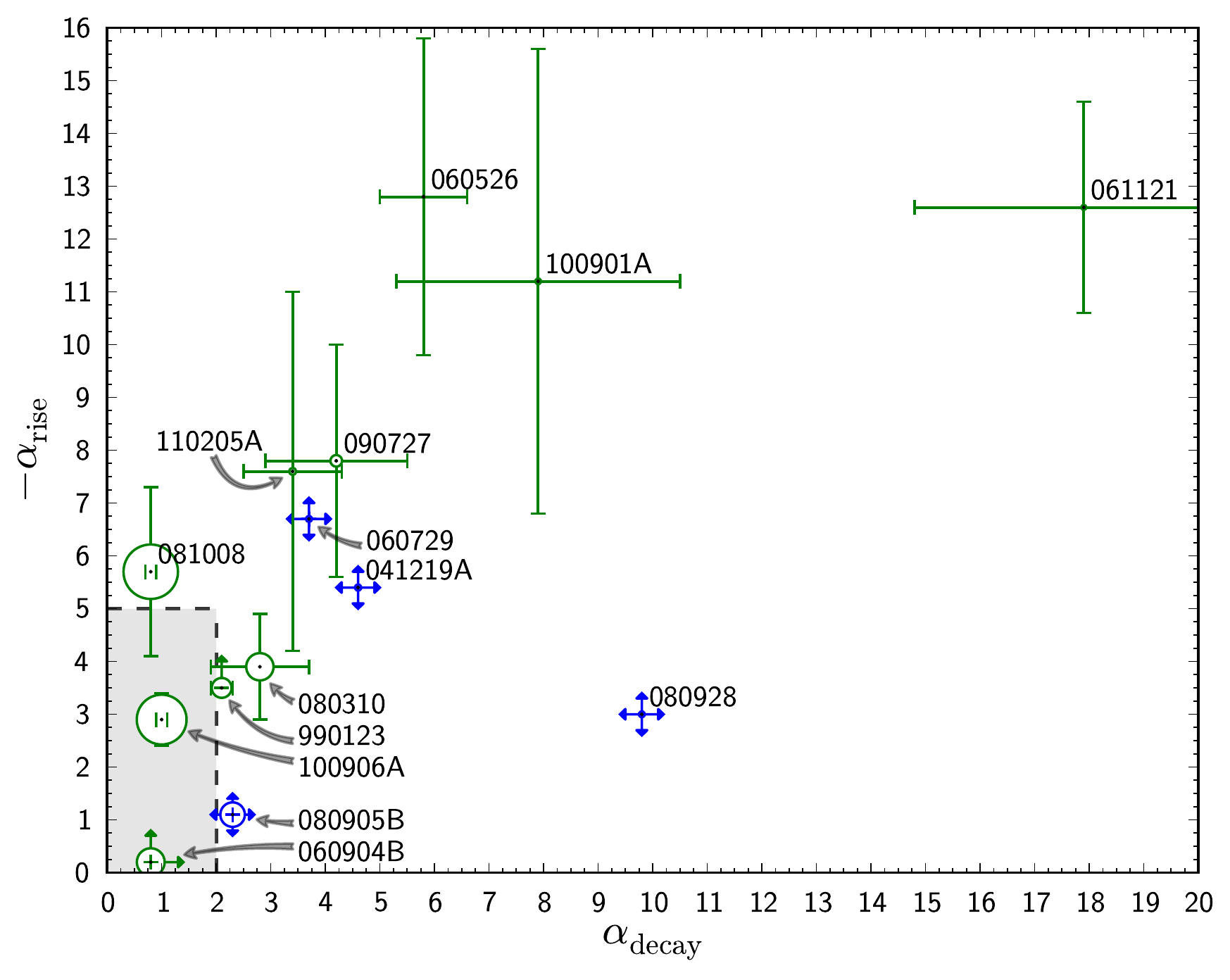}
\caption{\label{fig:risedecay} $|\alpha_\mathrm{rise}|$ versus $\alpha_\mathrm{decay}$ diagram for $14$ optical flashes for which both indices are given in Table \ref{tab:plindices}. Green points are obtained by fitting the optical LC or from the literature, while blue points represent rough estimates (as explained in text and in Appendix \ref{sect:aa} for each GRB). Gray shaded area shows the region of power-law indices as expected from the external shock model, i.e., $|\alpha_\mathrm{rise}| \lesssim 5$ and $\alpha_\mathrm{decay} \lesssim 2$ \citep{zhang2003}. The radius of the circle around each point represents the relative value of $\Delta t/t$ ($=w/t_\mathrm{peak}$) from Table \ref{tab:sampletaboptical} -- larger radius means higher $\Delta t/t$.}
\end{figure}

\par Steep rise and decay of early time optical emission (Figure \ref{fig:risedecay}) and sharp optical peaks ($\Delta t/t < 1$; see Figure \ref{fig:dtt}) likely imply the internal shock scenario for the majority of GRBs in the sub-sample. However, for a few cases (e.g., GRB~080810, GRB~081008, GRB~100906A) the large $\Delta t/t$ value from Table \ref{tab:sampletaboptical} (especially the values marked with the $^\dag$, which are not presented in Figure \ref{fig:dtt} and believably represent the domination of the afterglow emission) and low $\alpha_\mathrm{decay}$ and $|\alpha_\mathrm{rise}|$ indices are consistent with the external shock model, which could probably be the correct interpretation for these cases.

\subsection{Spectral properties}
\label{sect:specprop}
\par It is interesting to study prompt optical emission in the spectral domain. \citet{yost2007b}, for example, studied flux densities of optical and gamma-ray emission and found that gamma-ray spectrum may predict, overpredict or underpredict the optical flux, which leads to different interpretations. They concluded that there is no universal ratio between optical and gamma-ray flux densities. \citet{panaitescu2008} analysed $64$ GRBs with optical observations during prompt gamma-ray emission ($35$ among these have only an upper limit in the optical band) and found that for $10$ of them the optical flux is $1-4$ orders of magnitude brighter than the extrapolation of gamma-ray spectrum to optical frequencies. The diversity of the brightness of optical emission in GRB prompt phase is thus rather rich, but it can be biased towards bright optical emission. This is due to the fact that the smaller robotic telescopes usually start observations relatively quickly (e.g. tens of seconds) after the trigger, but are not able to detect faint emission (e.g. below $\sim 18\,\mathrm{mag}$).

\begin{figure*}[!ht]
\begin{center}
\includegraphics[width=0.8\linewidth]{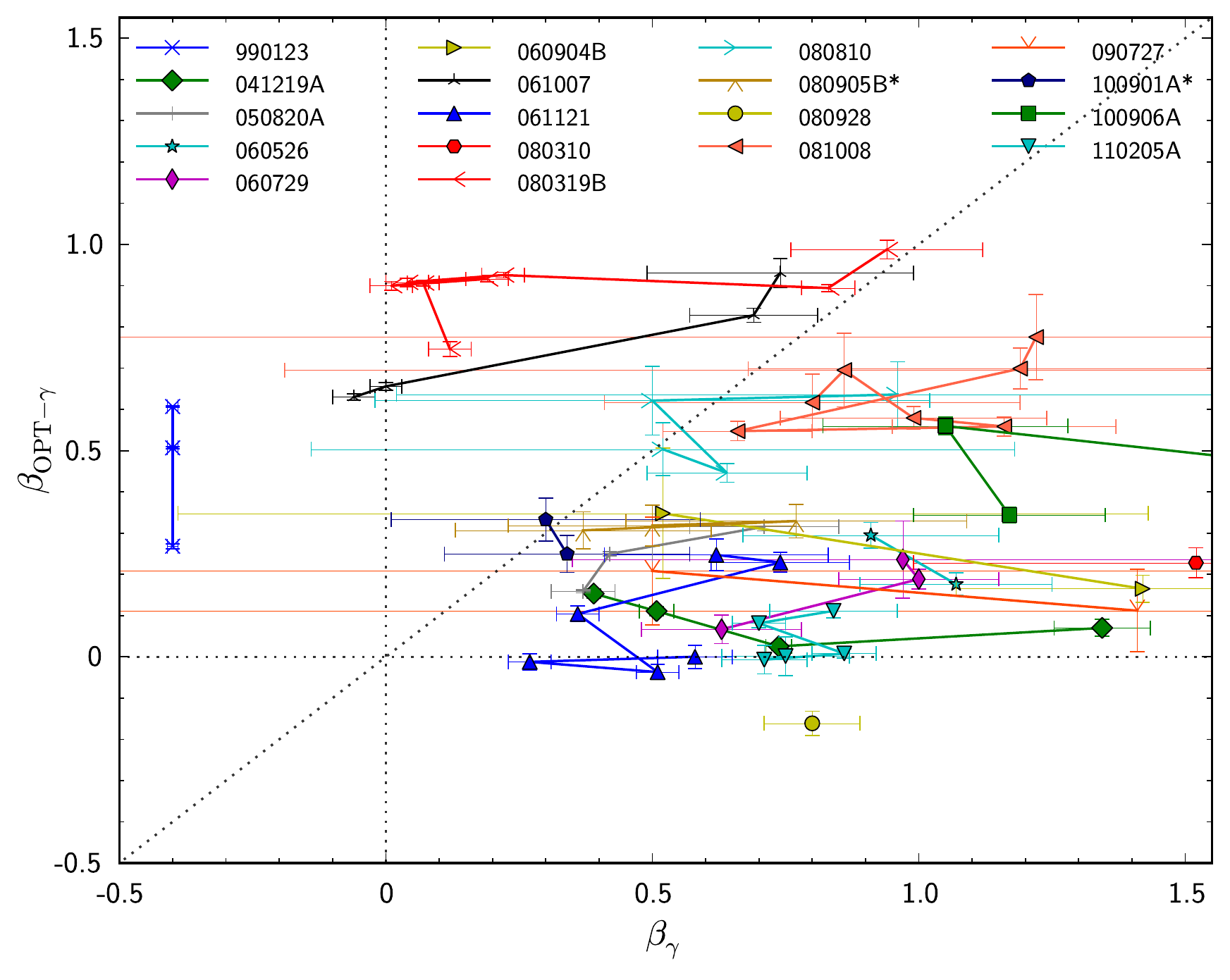}
\caption{\label{fig:beta} Optical--to--gamma-ray index ($\beta _{\mathrm{OPT}-\gamma}$) versus gamma-ray spectral index ($\beta _\gamma$) for GRBs in the sub-sample. Lines connect several optical detections during gamma-ray emission in a single GRB event. If $\beta _{\mathrm{OPT}-\gamma} < \beta _\gamma$, gamma-ray spectrum overpredicts the optical flux, indicating a spectral break (below diagonal line). If $\beta _{\mathrm{OPT}-\gamma} > \beta _\gamma$, gamma-ray spectrum underpredicts the optical flux, indicating a separate emission component in optical regime (above diagonal line).}
\end{center}
\end{figure*}

\par To study the spectral properties for \textit{Swift} GRBs in the sub-sample we have generated gamma-ray spectra in the time interval of interest (during the exposure time of the corresponding optical emission) and in $15-150\,\mathrm{keV}$ range using the \texttt{batbinevt} tool (we further used \texttt{batupdatephakw} and \texttt{batphasyserr} to apply corrections to the spectra, and we used the weighted count averaged energy response to fit the spectra if the time of interest included pre-slew, slew and post-slew intervals). We fitted the spectra with a simple power-law model and extracted the power-law indices $\Gamma$ ($= \beta_\gamma + 1$) and the flux densities at certain frequencies. We note that the BAT passband is relatively narrow and that $\beta_\gamma$ values are not so robust as they would be if using an instrument sensitive in a broader GRB spectral range, which typical extends to several hundreds $\mathrm{keV}$. For GRB~990123, GRB~041219A and GRB~050820A, the values were taken from \citet{yost2007a} and \citet{yost2007b}. We gathered optical LCs for the sub-sample from the literature (references are given in Appendices \ref{sect:a990123}-\ref{sect:a110205a}). We corrected the observed magnitudes for both the Galactic extinction \citep{schlegel1998} and the host galaxy extinction (where available, references are given in Appendix \ref{sect:aa}). The final values are gathered in Table \ref{tab:photomspec}.

\par In Figure \ref{fig:beta} we plot the power-law slope of gamma-ray spectrum ($\beta_\gamma$) versus the power-law slope of optical--to--gamma-ray index ($\beta_\mathrm{OPT-\gamma}$). This plot shows that gamma-ray spectrum can in principle overpredict (if $\beta _{\mathrm{OPT}-\gamma} < \beta _\gamma$) or underpredict (if $\beta _{\mathrm{OPT}-\gamma} > \beta _\gamma$) the optical flux. When optical flux is lower than the extrapolation of gamma-ray spectrum to the optical region, we could interpret it as an indication of a spectral break. This is the region below the diagonal line in Figure \ref{fig:beta}. On the other hand, when optical flux is higher than the extrapolation of gamma-ray spectrum to the optical region, this could be an indication that we are dealing with some additional, separate emission component in optical regime. This is the region above the diagonal line in Figure \ref{fig:beta}. In general, the inconsistency between $\beta_\gamma$ and $\beta_\mathrm{OPT-\gamma}$ could also be due to a different emission mechanisms.

\par Figure \ref{fig:beta} shows one case (GRB~080928) where $\beta _{\mathrm{OPT}-\gamma}$ is negative and one case (GRB~990123) where $\beta_\gamma$ is negative. In the first case, the behaviour could be explained by a steep spectral break or multiple spectral breaks in the SED, depending on the position of the synchrotron typical and cooling frequencies. The second case does definitely not represent a typical SED, but it could be explained in a context of distinct emission regions or different synchrotron processes in optical and gamma-ray regime. However, although both $\beta_\gamma$ and $\beta _{\mathrm{OPT}-\gamma}$ distributions in Figure \ref{fig:beta} show relatively large dispersions, they are within the range of $\beta = -1/3$ to $\beta = 3/2$, i.e., consistent with the synchrotron spectra of a relativistic shocks and electron power-law distribution indices of $p=2-3$ \citep{sari1998}.

\par In Figure \ref{fig:fratio} we show the temporal evolution of the flux ratios between gamma-ray and optical emission ($(\nu F_\nu)^\mathrm{gamma}/(\nu F_\nu)^\mathrm{opt}$), together with the corresponding histogram that shows the distribution of the average flux ratios for GRBs in the sub-sample (green histogram). Light blue normalised histogram in the background represents a Monte Carlo simulation that generates $10^4$ histograms by taking into account also the span of the flux ratios for every GRB. This information could otherwise be lost by averaging the flux ratio. The method is similar as described before by Eq. (\ref{eq:mc}).

\begin{figure*}[!ht]
\begin{center}
\includegraphics[width=0.8\linewidth]{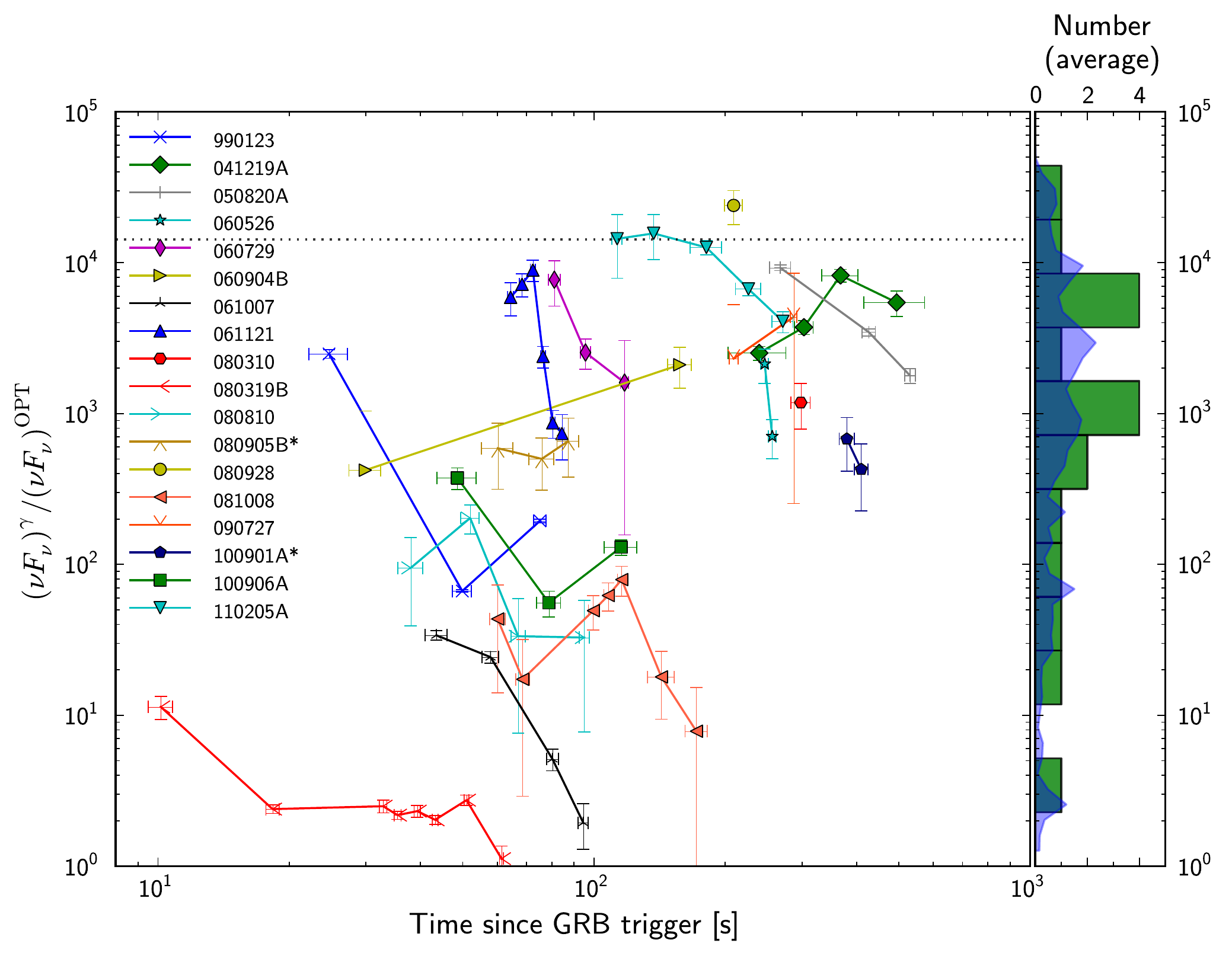}
\caption{\label{fig:fratio} Left: Temporal evolution of flux ratio between gamma-ray and optical emission for GRBs in the sub-sample. Region above the dotted line indicates $F_\nu^\gamma/F_\nu^\mathrm{OPT} > 1$ (using the average $\nu ^\gamma / \nu ^\mathrm{OPT} \approx 1.4\times 10^4$), corresponding to $\beta _{\mathrm{OPT}-\gamma} < 0$. Right: Green histogram represents the distribution of the average flux ratio. Light blue normalised histogram in the background represents a Monte Carlo simulation that takes into account also the span of the flux ratios for each GRB.}
\end{center}
\end{figure*}

\par We see that the $(\nu F_\nu)^\mathrm{gamma}/(\nu F_\nu)^\mathrm{opt}$ distribution spans over wide range of almost $5$ orders of magnitude, with no common flux ratio. This raises the question of whether it is possible for any single model to predict emission over such a wide range of flux ratios, or whether individual distinct models are necessary to explain prompt GRB emission case by case.

\section{Internal shock dissipation model}
\label{sect:simulation}
\par To test whether one model could explain early time optical emission occurring simultaneously with gamma-ray emission, we used a very simple internal shock dynamics model. The results from the observations -- namely sharp optical peaks and steep rise and decay of optical peaks -- hint at internal shock origin. Similarly, the diversity in the measured parameters implies some random processes, pointing towards prompt phase origin. Another advantage of this simple model is that the temporal correlation between gamma-ray and optical emission is not a requirement, as the synchrotron emission in different regimes does not necessary originate from the same location (same relativistic shells).

\subsection{Model}
\par Let us assume that the central engine of a GRB ejects a highly relativistic shell (A) with a certain bulk Lorentz factor $\gamma _\mathrm{A}$ and isotropic equivalent kinetic energy luminosity $L_\mathrm{A}$. Some time later ($\delta t$, which can be estimated from the variability of the light curve), another shell (B) is ejected with $\gamma _\mathrm{B}$ and $L_\mathrm{B}$. If $\gamma_\mathrm{B} > \gamma _\mathrm{A}$, the faster shell catches up with the slower one and the inelastic collision occurs. The collision radius for $(\gamma _\mathrm{A}, \gamma_\mathrm{B}) \gg 1$ is given by $R_\mathrm{is} = 2 \gamma_\mathrm{A}^2 c \delta t (1+z)^{-1}/(1-(\gamma_\mathrm{A}/\gamma _\mathrm{B})^2)$ \citep{yu2009}.

\par Due to the collision, a pair of internal shocks is generated: a forward and a reverse shock. Forward shock propagates into the shell A, while the reverse shock propagates into the shell B, and those are separated by the contact discontinuity surface. We denote the regions as: 1 = shell A, 2 = shocked shell A, 3 = shocked shell B, and 4 = shell B. Lorentz factors are $\gamma _1 = \gamma _\mathrm{A}$, $\gamma _4 = \gamma _\mathrm{B}$, $\gamma _2 = \gamma _3 \equiv \gamma$. Shocked regions 2 and 3 move relative to unshocked regions 1 and 4 with the following relative Lorentz factors \citep{yu2009}:
\begin{equation}
\label{eq:rellor}
\gamma_{21} = \frac{1}{2} \left( \frac{\gamma _1}{\gamma} + \frac{\gamma}{\gamma _1} \right), \quad \gamma_{34} = \frac{1}{2} \left(\frac{\gamma}{\gamma _4} + \frac{\gamma _4}{\gamma} \right)\,.
\end{equation}

\par According to the jump conditions between the two sides of a shock \citep{blandford1976}, the internal energy densities of the two shocked regions are $e_2 = (\gamma_{21}-1)(4\gamma_{21}+3)n_\mathrm{A} m_p c^2$ and $e_3 = (\gamma_{34}-1)(4\gamma_{34}+3)n_\mathrm{B} m_p c^2$, where $n_\mathrm{(A,B)} = L_\mathrm{(A,B)}/(4\pi R_\mathrm{is}^2 \gamma _\mathrm{(A,B)} ^2 m_p c^3)$. From the mechanical equilibrium requirement ($e_2 = e_3$), we get:
\begin{equation}
\label{eq:equilibr}
\frac{(\gamma_{21}-1)(4\gamma_{21}+3)}{(\gamma_{34}-1)(4\gamma_{34}+3)} = \frac{n_\mathrm{B}}{n_\mathrm{A}} = \frac{L_\mathrm{B}}{L_\mathrm{A}} \left(\frac{\gamma _1}{\gamma _4} \right)^2\,.
\end{equation}
By solving the Equations (\ref{eq:rellor}) and (\ref{eq:equilibr}) we obtain the value of $\gamma$, which is the Lorentz factor of the shocked regions (the system usually has more than one solution, but only one of them is real and matches the criterion $\gamma _\mathrm{A} < \gamma < \gamma _\mathrm{B}$). 

\par Now that we have $\gamma$, $\gamma_{21}$ and $\gamma_{34}$ (which are constant during shock crossing the shell, \citealt{yu2009}), we can obtain the synchrotron spectrum of relativistic electrons in magnetic field. Following \citet{dai2002} and \citet{yu2009} we calculate the total number of electrons in the forward ($N_{e,2}$) and reverse ($N_{e,3}$) shocked regions, the strength of the magnetic fields ($B_2$, $B_3$), the minimum ($\gamma_{e,m,2}$, $\gamma_{e,m,3}$) and the cooling ($\gamma_{e,c,2}$, $\gamma_{e,c,3}$) Lorentz factors, and finally the characteristic synchrotron frequencies and peak flux densities \citep{sari1998, fanwei2005}:
\begin{eqnarray}
& \nu _{m,(2,3)} = \frac{q_e}{2 \pi m_e c (1+z)} B_{(2,3)} \gamma \gamma _{e,m,(2,3)}^2 \,, \notag \\
& \nu _{c,(2,3)} = \frac{q_e}{2 \pi m_e c (1+z)} B_{(2,3)} \gamma \gamma _{e,c,(2,3)}^2 \,, \notag \\
& F_{\nu,\mathrm{max},(2,3)} \approx \frac{3 \sqrt{3} \,0.6\, (1+z) m_e c^2 \sigma _T N_{e,(2,3)}}{32 \pi^2 q_e d_L^2} B_{(2,3)} \gamma \,.
\end{eqnarray}
Using the synchrotron spectrum formulae \citep{sari1998} we obtain:
\begin{equation}
\label{eq:syncflux}
\begin{array}{c}
F_{\nu,(2,3)} = \\ \\
F_{\nu,\max,(2,3)}\times \left\{
\begin{array}{ll}
\left({\nu\over\nu_l}\right)^{1/3},~~~~~~~~~~~~~\hfill \nu<\nu_l\\
\left({\nu\over\nu_l}\right)^{-(q-1)/2},~~~~~~~~~~~~~\hfill \nu_l<\nu<\nu_h\\
\left({\nu_h\over\nu_l}\right)^{-(q-1)/2}\left({\nu\over\nu_h}\right)^{-p/2},~~~~~~~~~~~~~\hfill \nu_h<\nu\\
\end{array}\right.
\end{array}
\end{equation}
where 
\begin{center}
$\nu_{l}=\min(\nu_{m,(2,3)},\nu_{c,(2,3)})\,,$\\
$\nu_{h}=\max(\nu_{m,(2,3)},\nu_{c,(2,3)})\,,$
\end{center}
and $q=2$ for $\nu_{c,(2,3)}<\nu_{m,(2,3)}$ and $q=p$ for $\nu_{c,(2,3)}>\nu_{m,(2,3)}$. By replacing $\nu$ with the typical gamma-ray and optical frequencies, we can calculate the flux density of gamma-ray and optical emission from the synchrotron model in the forward and reverse shock regions.
 
\par Finally, we draw attention to the synchrotron self-absorption (SSA), which is able to suppress gamma-ray or optical emission \citep{panaitescukumar2000, yu2009, shen2009}. We calculated the SSA thickness ($\tau _\mathrm{SSA}$) at gamma-ray and optical frequencies and ignored the SSA break if $\tau _\mathrm{SSA}$ is much smaller than $1$.

\subsection{Simulation and results}
\par Following the method described above we generated $100$ events (GRBs), among which every event produced $50$ relativistic forward and reverse shocks that emitted synchrotron radiation as described by Eq. (\ref{eq:syncflux}). For $\gamma _A$ and $\gamma _B$ we chose random values from $\gamma _{\min} = 50$ to $\gamma _{\max} = 10000$ that were uniformly distributed in log space, and required $\gamma _B > 100 \times \gamma _A$ to increase the efficiency of the internal shocks \citep{kobayashi1997}. Similarly, we chose random values for the luminosities $L_A$ and $L_B$ that were uniformly distributed in log space from $L_{\min} = 10^{46} \, \mathrm{erg \; s^{-1}}$ to $L_{\max} = 10^{50} \, \mathrm{erg \; s^{-1}}$, without specifying which luminosity should be higher. For the variability time-scale $\delta t$ we chose random values from $\delta t _{\min}= 0.01\,\mathrm{s}$ to $\delta t _{\max}= 10\,\mathrm{s}$, uniformly distributed in linear space. We randomly chose the redshift in the interval from $z_{\min} = 0.5$ to $z_{\max} = 3.5$ (roughly consistent with the sample) and calculated the corresponding luminosity distance ($d_L$).

\par We assumed $p=2.5$ for power-law electron distribution, and chose $\nu ^\mathrm{OPT} = 5.4 \times 10^{14} \,\mathrm{Hz}$ and $\nu ^\gamma = 7.8 \times 10^{18} \,\mathrm{Hz}$, as obtained from the observations. We used $\epsilon _e = 0.5$ and $\epsilon _B=0.01$. With these initial parameters we run the calculation described in the previous section.

\par In Figure \ref{fig:fratio_simulation} we present the results obtained from the simulation, in the form of normalised flux ratio distribution for the sum of both the forward and the reverse internal shocks contributions: 
\begin{equation}
\frac{\left((\nu F_\nu)_\mathrm{FS} + (\nu F_\nu)_\mathrm{RS}\right)^\gamma}{\left((\nu F_\nu)_\mathrm{FS} + (\nu F_\nu)_\mathrm{RS}\right)^\mathrm{OPT}} \,.
\end{equation}
The height of the distribution obtained from the simulation depends on the normalisation and does not give any direct physical information. However, when comparing the results from the simulation and the observed flux ratio distribution (from Figure \ref{fig:fratio}) it is evident that the synchrotron emission during internal shocks can produce radiation spanning over a wide range of flux ratios. By examining the parameter space using a simple internal shock dissipation model, we were able to obtain comparable results to those obtained from the observations. The only requirement that we used was that the two colliding shells have very different Lorentz factors.

\begin{figure}[!h]
\begin{center}
\includegraphics[width=1\linewidth]{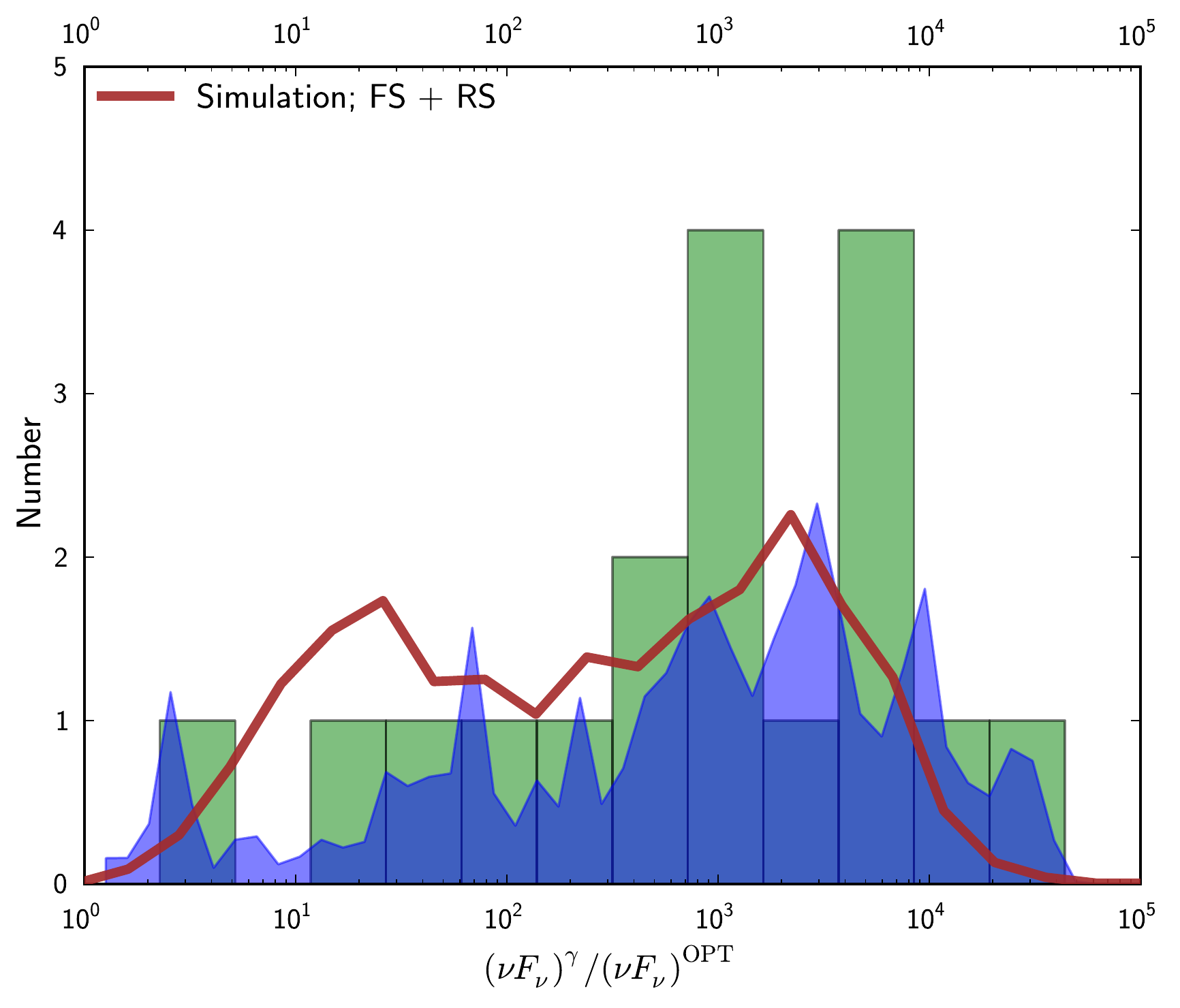} 
\caption{\label{fig:fratio_simulation} Flux ratio distributions from observations (green and blue histograms, see Figure \ref{fig:fratio}) and from the internal shock dissipation model simulation (normalised brown solid line for the sum of both forward and reverse internal shocks contributions). }
\end{center}
\end{figure}

\section{Conclusion and Discussion}
\label{sect:conclusion}

\par We presented a detailed analysis of GRB~090727, which shows an early time optical emission during the prompt gamma-ray emission. By modelling the early time optical light curve as obtained from the Liverpool Telescope and comparing the peak time and the shape of the optical flash with the contemporaneous X-ray peak, the most probable scenario is that the peaks are temporally non-simultaneous, although this is not conclusive due to the scarcity of observational data points during the peak time.

\par The rise ($\alpha _\mathrm{rise} \approx -7.8 \pm 2.2$) and decay ($\alpha _\mathrm{decay} \approx 4.2 \pm 1.3$) indices for the early time optical peak, as well as the spectral evolution and rather low X-ray to optical spectral index, disfavour the early time optical peak as an afterglow emission from the external shock, either from the forward-external or reverse-external shock. Instead, we conclude that the early time optical flash likely originates from an internal shock region. If the early time optical and X-ray peaks are simultaneous, this would suggest that they are produced by the same internal shock, whereas if they are non-simultaneous, the peaks originate from a slightly different internal shock (or other internal dissipation mechanism) components.

\par To study GRB early time optical emission in more detail, we select a sub-sample of $18$ GRBs for which optical emission peaks during gamma-ray emission. We study early time temporal and spectral properties. We fitted optical LCs with Norris and Gaussian profiles and determined the peak time and the peak duration. We obtained the rise and decay power-law indices of early time optical peaks from the literature or by fitting the Beuermann profile or a simple power-law behaviour. We calculated gamma-ray spectral indices, optical--to--gamma-ray spectral indices and flux ratios between optical and gamma-ray. 

\par We found a large diversity in all these observed properties, which strongly imply some random processes. The lack of any common features thus challenge the theoretical models for prompt GRB emission. Furthermore, we found that the clear correlation (same peak time and power-law indices) between gamma-ray and optical emission during prompt GRB phase is rare, but this could also be due to the sparse optical coverage.

\par Using a simple internal shock dissipation model, in which relativistic shells collide and form two internal shocks (forward-internal and reverse-internal), we examine the parameter space. By using random initial parameters from a typical GRB parameter space we simulate the evolution of relativistic shocks. The results show that dissipation within internal shocks can produce radiation at optical and gamma-ray frequencies and that the flux ratios between optical and gamma-ray can be diverse, spanning over many orders of magnitude. This is in agreement with the observations. We therefore conclude that the internal shocks could be the origin of both gamma-ray and early time optical flashes for a majority of GRBs in the sub-sample.

\par For a few GRBs, however, we can not exclude other mechanisms, especially where properties like power-law indices, $\Delta t/t$ or $\beta _{\mathrm{OPT}-\gamma}$ values point towards certain theoretical description that is perhaps independent of the optical to gamma-ray flux ratio. The optical flash in GRB~990123 is believed to be produced by the reverse-external shock (\ref{sect:a990123}); optical light curves in GRB~080810 and GRB~081008 are likely produced by the forward-external shock, but superimposed optical flashes may be due to internal shocks (\ref{sect:a080810},\ref{sect:a081008}); early time optical emission in GRB~080310 is probably due to reverse-external shock, based on the power-law indices and $\Delta t/t$; GRB~100906A shows non-sharp and shallow optical light curve, which is indicative of the forward-external shock origin. All these GRBs show rather shallow decay/rise indices and rather large $\Delta t/t$ (see Figure \ref{fig:risedecay}). For other GRBs in the sub-sample we propose dissipation within internal shocks as an origin for early time optical flashes.

\acknowledgements{
This work made use of data supplied by the UK Swift Science Data Centre at the University of Leicester. We acknowledge support from STFC and the Royal Society. Swift mission is funded in the UK by STFC, in Italy by ASI, and in the USA by NASA. The Faulkes Telescopes are operated by the Las Cumbres Observatory. The Liverpool Telescope is owned and operated by Liverpool John Moores University. A.G. acknowledges funding from the Slovenian Research Agency and from the Centre of Excellence for Space Sciences and Technologies SPACE-SI, an operation partly financed by the European Union, European Regional Development Fund and Republic of Slovenia, Ministry of Higher Education, Science and Technology.}

\begin{appendix}

\section{GRBs with early time optical emission in the sub-sample}
\label{sect:aa}
\par In the following subsections we summarise the basic properties of $17$ GRBs with early time optical emission (besides GRB~090727) from the sub-sample.

\subsection{GRB~990123}
\label{sect:a990123}
\par \textbf{Literature}: Gamma-ray emission was due to inverse Comptonisation, while the simultaneous optical emission was produced by synchrotron processes, either in the internal or the reverse-external shocks \citep{panaitescukumar2007}. \citet{vestrand2005} suggested that the optical emission was generated by different process than the prompt gamma-ray emission, i.e., by the reverse external shock arising from the ejecta's collision with the surrounding material.
\par \textbf{Energetics}: At redshift $z=1.6$ this GRB had an isotropic gamma-ray energy release of $E_{\gamma,\mathrm{iso}} = 1.6 \times 10^{54}\,\mathrm{ergs}$ above $20\,\mathrm{keV}$ \citep{briggs1999}. 
\par \textbf{Light curve}: Power-law indices for the optical peak are $\alpha _\mathrm{rise} < -3.5$ and $\alpha _\mathrm{decay} = 2.1 \pm 0.2$ \citep{melandri2010}.
\par Optical data in V equivalent band were obtained from \citet{akerlof1999}. They calibrated the magnitudes against the Tycho catalogue. To fit the Norris profile to the early time optical peak, we added $\pm 0.15\,\mathrm{mag}$ systematic error in quadrature to the original photometric errors, as inferred from \citet{akerlof1999}.

\subsection{GRB~041219A}
\label{sect:a041219a}
\par \textbf{Literature}: Emission at different frequencies is likely coming from single emission region. Optical component was due to internal shocks in the GRB outflow. The correlation between optical and gamma-ray emission suggests a common mechanism \citep{vestrand2005}. There are some alternative possible emission mechanisms, like for example saturated Comptonization \citep{vestrand2005, zheng2006}.
\par Initial optical peak is followed by a short lasting and powerful infra-red flash at $\sim 450\,\mathrm{s}$ after the trigger \citep{blake2005}. \citet{fan2005} interpreted that as an external reverse shock, superimposed on forward shock afterglow emission. This interpretation is also supported by the radio data \citep{fan2005}.
\par \textbf{Energetics}: There was no redshift measured for this GRB, but the best fit solution gives $z=0.31^{+0.54}_{-0.26}$ for the photometric redshift \citep{goetz2011}. This corresponds to an isotropic gamma-ray energy release of $2 \times 10^{51} \,\mathrm{ergs} \leq E_{\gamma,\mathrm{iso}} \leq 2 \times 10^{54}\,\mathrm{ergs}$ in $1\,\mathrm{keV} - 10\,\mathrm{MeV}$ band \citep{goetz2011}.
\par \textbf{Light curve}: During the optical peak, the optical coverage is insufficient to perform the fit, but if we take the first two and the last optical data point from the peak (Figure \ref{fig:samplelcs}) and simply estimate the power-law indices we get roughly $\alpha _\mathrm{rise} \sim -5.4$ and $\alpha _\mathrm{decay} \sim 4.6$.
\par Optical data in $\mathrm{R_C}$ band were obtained from \citet{vestrand2005}.

\subsection{GRB~050820A}
\label{sect:a050820a}
\par \textbf{Literature}: The early time optical LC shows a complex shape due to contributions from internal and external shocks. The onset of the afterglow occurs during the prompt GRB phase. The weak optical flash before the onset of the afterglow is likely associated with the internal shocks \citep{cenko2006, genet2007}.
\par \textbf{Energetics}: At redshift $z=2.615$ this GRB had an isotropic gamma-ray energy release of $E_{\gamma,\mathrm{iso}} = 8.3^{+2.5}_{-1.1} \times 10^{53}\,\mathrm{ergs}$ in $0.02 - 2\,\mathrm{MeV}$ band \citep{cenko2006}.
\par \textbf{Light curve}: The LC is complex, so it is not obvious which rise or decay index is consistent with the observed evolution. The optical afterglow emission peaks at $\sim 425\,\mathrm{s}$ after the trigger, when gamma-ray emission is still active (Figure \ref{fig:samplelcs}). After that the optical afterglow decays with $\alpha _\mathrm{decay} = 0.97 \pm 0.01$ \citep{melandri2010}.
\par Optical data in R-equivalent band were obtained from \citet{vestrand2006}. They calibrated the magnitudes against the USNO-B1 catalogue. To fit the Norris profile to the early time optical peak, we added $\pm 0.2\,\mathrm{mag}$ systematic error in quadrature to the original photometric errors.

\subsection{GRB~060526}
\label{sect:a060526}
\par \textbf{Literature}: \citet{thoene2010} argued that the central engine activity dominates the early time optical light curve. Early time optical emission could be associated with the late internal shock scenario. The optical LC at later times shows some deviations from the smooth power-law behaviour, which can be explained by energy injections from the central engine \citep{dai2007, thoene2010}.
\par \textbf{Energetics}: At redshift $z=3.221$ this GRB had an isotropic gamma-ray energy release of $E_{\gamma,\mathrm{iso}} \approx 2.4 \times 10^{52}\,\mathrm{ergs}$ in $15-150\,\mathrm{keV}$ band \citep{thoene2010}.
\par \textbf{Light curve}: Rise and decay power-law indices for the optical flash which peaks at $\sim 256\,\mathrm{s}$ after the trigger are $\alpha _\mathrm{rise} = -12.8 \pm 3.0$ and $\alpha _\mathrm{decay} = 5.8 \pm 0.8$ \citep{thoene2010}.
\par Optical data in the UVOT v band were obtained from \citet{thoene2010}. For the purpose of the spectral analysis we corrected the magnitudes for additional host galaxy absorption of $A_\mathrm{V}^\mathrm{rest} = 0.06$, or, assuming the SMC-like extinction, for $A_\mathrm{v} = 0.26 \,\mathrm{mag}$ in the observer frame \citep{thoene2010}.

\subsection{GRB~060729}
\label{sect:a060729}
\par \textbf{Literature}: No particular model for the early time optical peak has been proposed in the literature. The first optical peak is extremely dim with respect to the gamma-ray or X-ray brightness \citep{rykoff2009}. The very long-lived and extremely flat optical afterglow LC at later times could be due to the long-duration energy injection \citep{grupe2007, xu2009}.
\par \textbf{Energetics}: At redshift $z=0.54$ this GRB had an isotropic gamma-ray energy release of $E_{\gamma,\mathrm{iso}} \leq 1.6 \times 10^{52}\,\mathrm{ergs}$ in $1\,\mathrm{keV}-10\,\mathrm{MeV}$ band \citep{grupe2007}.
\par \textbf{Light curve}: During the first optical peak at $95\,\mathrm{s}$ after the trigger the coverage is insufficient to perform the fit \citep{rykoff2009}. If we take the first two and the fifth optical data points (Figure \ref{fig:samplelcs}) and simply estimate the power-law indices we get roughly $\alpha _\mathrm{rise} \sim -6.7$ and $\alpha _\mathrm{decay} \sim 3.7$. 
\par Optical data in $\mathrm{R_C}$ band were obtained from \citet{rykoff2009}. For the purpose of the spectral analysis we corrected the magnitudes for additional host galaxy absorption of $A_\mathrm{V}^\mathrm{rest} = 0.07$, or, assuming the SMC-like extinction, for $A_\mathrm{R_C} = 0.1 \,\mathrm{mag}$ in the observer frame \citep{zafar2011}.

\subsection{GRB~060904B}
\label{sect:a060904b}
\par \textbf{Literature}: This GRB shows similar behaviour as GRB~060729. The X-ray flare at $\sim 170\,\mathrm{s}$ after the trigger is very bright and comparable with the optical emission at that time. The optical LC is extremely flat at later times \citep{rykoff2009}. \citet{klotz2008} argued that the optical peak coincides with the initial X-ray flare, but due to different temporal decay indices it is not obvious if peaks are correlated and if the emission is coming from single region. 
\par \textbf{Energetics}: At redshift $z=0.703$ this GRB had an isotropic gamma-ray energy release of $E_{\gamma,\mathrm{iso}} \sim 2 \times 10^{51} \,\mathrm{ergs}$ in $15-150\,\mathrm{keV}$ band \citep{klotz2008}.
\par \textbf{Light curve}: Early time optical LC shows many peaks. The first peak occurs at $\sim 55\,\mathrm{s}$ after the trigger and the limits for the power-law indices are $\alpha _\mathrm{rise} < -0.8$ and $\alpha _\mathrm{decay} > 0.2$ \citep{rykoff2009}. Our rough estimation gives $\alpha _\mathrm{rise} \sim -1.1$ and $\alpha _\mathrm{decay} \sim 1.3$. After the end of the first peak the optical LC shows complex behaviour until $\sim 585\,\mathrm{s}$ after the trigger, when the optical emission reaches the maximum brightness and then decays with the power-law index $\alpha _\mathrm{decay} \approx 1$. At $\sim 1000\,\mathrm{s}$ after the trigger the optical plateau occurs, lasting for about $1500\,\mathrm{s}$. The emission is again decaying with the power law index $\alpha _\mathrm{decay} \approx 1$ after the end of the plateau. 
\par Optical data in $\mathrm{R_C}$ band were obtained from \citet{rykoff2009}. For the purpose of the spectral analysis we corrected the magnitudes for additional host galaxy absorption of $A_\mathrm{V}^\mathrm{rest} = 0.08$, or, assuming the SMC-like extinction, for $A_\mathrm{R_C} = 0.12 \,\mathrm{mag}$ in the observer frame \citep{kann2010}.

\subsection{GRB~061007}
\label{sect:a061007}
\par \textbf{Literature}: While the late time optical LC is consistent with the forward-external shock afterglow emission \citep{mundell2007b}, \citet{rykoff2009} argued that perhaps at early times some prompt GRB processes contribute to the emission in the optical regime.
\par \textbf{Energetics}: At redshift $z=1.261$ this GRB had an isotropic gamma-ray energy release of $E_{\gamma,\mathrm{iso}} \sim 10^{54} \,\mathrm{ergs}$ in $1\,\mathrm{keV}-10\,\mathrm{MeV}$ band \citep{mundell2007b}.
\par \textbf{Light curve}: It is not clear weather the early time optical peak occurs at $\approx 58\,\mathrm{s}$ or at $\approx 95\,\mathrm{s}$ after the trigger and if there is only one or maybe two peaks. Reported power-law indices are $\alpha _\mathrm{rise} \sim -9$ and $\alpha _\mathrm{decay} = 1.7 \pm 0.1$ \citep{melandri2010}.
\par Optical data in $\mathrm{R_C}$ band were obtained from \citet{rykoff2009}. For the purpose of the spectral analysis we corrected the magnitudes for additional host galaxy absorption of $A_\mathrm{V}^\mathrm{rest} = 0.48$, or, assuming the SMC-like extinction, for $A_\mathrm{R_C} = 0.98 \,\mathrm{mag}$ in the observer frame \citep{mundell2007b}.

\subsection{GRB~061121}
\label{sect:a061121}
\par \textbf{Literature}: The SED during the prompt emission is harder than predicted by the standard fireball model, but spectral curvature around the break could flatten the spectral index \citep{page2007}. \citet{giannios2008} suggested that the emission could be explained by the Poynting-flux-dominated outflow where magnetic reconnection gradually dissipates the energy. This mechanism predicts the prompt optical emission component associated with the gamma-ray emission.
\par \textbf{Energetics}: At redshift $z=1.314$ this GRB had an isotropic gamma-ray energy release of $E_{\gamma,\mathrm{iso}} = 2.8\times 10^{53} \,\mathrm{ergs}$ in $1\,\mathrm{keV}-10\,\mathrm{MeV}$ band \citep{page2007}.
\par \textbf{Light curve}: By fitting the Beuermann profile (Eq. \ref{eq:beuermann}) to the early time optical peak in the time interval $61-92\,\mathrm{s}$ after the trigger, we get $\alpha _\mathrm{rise} = -12.6 \pm 2.0$ and $\alpha _\mathrm{decay} = 17.9 \pm 3.1$ ($\chi _\mathrm{red} ^2$ is $0.8$ with $11$ d.o.f.). At later times the optical LC decays differently as the X-ray LC until $\sim 50\,\mathrm{ks}$ after the trigger \citep{uehara2011}. Such LCs decoupling could be achieved by variations of the microphysical parameters or by an incoming outflow from the central engine, adding energy to the blast wave \citep{panaitescuvestrand2011, uehara2011}.
\par Optical data in the UVOT white band were obtained from \citet{page2007}. For the purpose of the spectral analysis we corrected the magnitudes for additional host galaxy absorption of $A_\mathrm{V}^\mathrm{rest} = 0.28$, or, assuming the MW-like extinction, for $A_\mathrm{w} = 1 \,\mathrm{mag}$ in the observer frame \citep{page2007}.

\subsection{GRB~080310}
\label{sect:a080310}
\par \textbf{Literature}: The optical LC is quite complex. It likely shows two bumps, with the first one showing fast rise and occurring simultaneously with the gamma-ray emission \citep{littlejohns2012}. Both bumps are relatively long-lasting and look like a combination of the reverse-external and the forward-external shock afterglow emission (Type I from \citealt{gomboc2009}), with some prompt component likely affecting the early time optical behaviour \citep{littlejohns2012}.
\par \textbf{Energetics}: At redshift $z = 2.43$ this GRB had an isotropic gamma-ray energy release of $E_{\gamma,\mathrm{iso}} = (3.2 \pm 0.3)\times 10^{52} \,\mathrm{ergs}$ in $15-150\,\mathrm{keV}$ band \citep{littlejohns2012}.
\par \textbf{Light curve}: By fitting the sum of two Beuermann profiles (Eq. \ref{eq:beuermann}) to the early time optical LC in the time interval $161-3100\,\mathrm{s}$ after the trigger, we get for the first bump $\alpha _\mathrm{rise} = -3.9 \pm 1.0$ and $\alpha _\mathrm{decay} = 2.8 \pm 0.9$, while for the second bump we get $\alpha _\mathrm{rise} = -1.0 \pm 0.5$ and $\alpha _\mathrm{decay} = 1.0 \pm 0.4$ ($\chi _\mathrm{red} ^2$ is $0.9$ with $14$ d.o.f.). From this analysis we found that the first bump peaks at $t_\mathrm{peak} = 255 \pm 14 \,\mathrm{s}$ after the trigger, while the second bump peaks at $t_\mathrm{peak} = 1337 \pm 553 \,\mathrm{s}$ after the trigger. At later times ($t \gtrsim 10^4\,\mathrm{s}$) the optical LC decays with the power-law index $\alpha _\mathrm{decay} \approx 1.3$ \citep{littlejohns2012}. At early times (during the first optical bump), the optical LC is likely influenced by some contribution from prompt processes, which results in rapid optical flux variations especially in the UVOT white band. 
\par Optical data in R band (we used first two data points from the Super-LOTIS telescope and other data points from the P60 and the VLT telescopes; see \citealt{littlejohns2012} for details) were obtained from \citet{littlejohns2012}. For the purpose of the spectral analysis we corrected the magnitudes for additional host galaxy absorption of $A_\mathrm{V}^\mathrm{rest} = 0.19$, or, assuming the SMC-like extinction, for $A_\mathrm{R} = 0.61 \,\mathrm{mag}$ in the observer frame \citep{littlejohns2012}.

\subsection{GRB~080319B}
\label{sect:a080319b}
\par \textbf{Literature}: Extreme optical flux suggests that the emission could be produced by the synchrotron self-Compton (SSC) process \citep{wozniak2009}. Degree of correlation between optical and gamma-ray emission can be very sensitive to several factors, among which the important one is the location of the synchrotron peak, which should be positioned in the optical band in this case \citep{wozniak2009}. \citet{racusin2008} proposed a two-component jet model together with the SSC process (which is not the only viable candidate) to explain high flux density of the optical flash during the prompt phase and the shapes of the LCs during prompt and afterglow phases. On the other hand, as the SSC model would imply additional spectral component arising from the second-order up-scattering to the GeV range, \citet{zou2009} concluded that for GRB~080319B the SSC model is not correct interpretation and that the gamma-ray and optical emission are in fact not coming from the single emission region (see also \citealt{zou2009b}). \citet{yuwangdai2009} proposed that optical and gamma-ray emission could both be produced by the internal shocks via synchrotron emission if the Lorentz factors of relativistic shells have a bimodal distribution.
\par \textbf{Energetics}: At redshift $z=0.937$ this GRB had an isotropic gamma-ray energy release of $E_{\gamma,\mathrm{iso}} = 1.3 \times 10^{54} \,\mathrm{ergs}$ in $20\,\mathrm{keV} - 7 \,\mathrm{MeV}$ band \citep{racusin2008}.
\par \textbf{Light curve}: The early time gamma-ray LC is composed of multiple peaks. Similarly, the optical LC is complex and shows the presence of $4$ early time peaks, which are nearly equidistant \citep{beskin2010}. Optical emission rises until around $\approx 20\,\mathrm{s}$ after the trigger, then it decays slightly but rises again at around $\approx 45\,\mathrm{s}$ after the trigger. Reported rise and decay power-law indices that correspond to the rise of the first peak and the decay of the last peak are $\alpha _\mathrm{rise} = -4.6 \pm 0.7$ \citep{pandey2009} and $\alpha _\mathrm{decay} = 6.5 \pm 0.9$ \citep{racusin2008}, respectively.
\par V-band optical data were obtained from \citet{racusin2008}. For the purpose of the spectral analysis we corrected the magnitudes for additional host galaxy absorption of $A_\mathrm{V}^\mathrm{rest} = 0.07$, or, assuming the SMC-like extinction, for $A_\mathrm{V} = 0.14 \,\mathrm{mag}$ in the observer frame \citep{bloom2009}.

\subsection{GRB~080810}
\label{sect:a080810}
\par \textbf{Literature}: Optical emission is probably the afterglow emission from the forward-external shock, occurring at very early times \citep{page2009}. 
\par \textbf{Energetics}: At redshift $z=3.355$ this GRB had an isotropic gamma-ray energy release of $E_{\gamma,\mathrm{iso}} \sim 3 \times 10^{53} \,\mathrm{ergs}$ in $1\,\mathrm{keV} - 10 \,\mathrm{MeV}$ band \citep{page2009}.
\par This GRB was also detected by the \textit{Fermi} GBM \citep{meegan2008}. Spectral analysis of the GBM data shows that the spectrum is best fitted with the power-law function with exponential high-energy cut-off. Values of the parameters of the fit are: power-law index $\beta = -1.13 \pm 0.05$, cut-off energy $E_\mathrm{peak} = 835 \pm 160\,\mathrm{keV}$ and fluence ($8-1000\,\mathrm{keV}$) $f = (1.25 \pm 0.03)\times 10^{-5}\,\mathrm{erg/cm^2}$. The angle from the \textit{Fermi} LAT boresight was $61^\circ$ \citep{bissaldi2009}.
\par \textbf{Light curve}: Reported power-law indices during the early time optical emission are $\alpha _\mathrm{rise} = -1.32 \pm 0.11$ and $\alpha _\mathrm{decay} = 1.22 \pm 0.09$ \citep{page2009,melandri2010}. The decay index is also consistent with the late time data. In addition to the optical afterglow LC, there is a non-prominent optical flare present at $\approx 100\,\mathrm{s}$ after the trigger, which happens simultaneous with the gamma-ray/X-ray flare at that time (Figure \ref{fig:samplelcs}). However, we are unable to determine the power-law indices while there are only a few optical data points.
\par Optical data in R-equivalent band were obtained from \citet{page2009}.

\subsection{GRB~080905B}
\label{sect:a080905b}
\par \textbf{Literature}: No particular model for the early time optical peak was proposed in the literature. 
\par \textbf{Energetics}: At redshift $z = 2.374$ \citep{vreeswijk2008} this GRB had an isotropic gamma-ray energy release of $E_{\gamma,\mathrm{iso}} =2.4 \times 10^{52} \,\mathrm{ergs}$ in $15-150\,\mathrm{keV}$ band (using the fluence reported in \citealt{barthelmy2008}).
\par This GRB was also detected by the \textit{Fermi} GBM \citep{bhat2008}. Similarly as the \textit{Swift} BAT, GBM detected the main peak at the BAT trigger time, followed by the second peak at $t \approx 70\,\mathrm{s}$ and the third peak lasting from $t \approx 80\,\mathrm{s}$ to $t \approx 110\,\mathrm{s}$. Preliminary results of the spectral analysis of the GBM data show that the time-averaged spectrum is well fitted with the power-law function with power-law index $\beta = -1.75 \pm 0.12$ and fluence ($20-1000\,\mathrm{keV}$) $f = (4.1 \pm 0.3)\times 10^{-8}\,\mathrm{erg/cm^2}$. The angle from the \textit{Fermi} LAT boresight was $82^\circ$ \citep{bhat2008}.
\par \textbf{Light curve}: The results are preliminary (marked with $^\ast$ in Figures \ref{fig:lc_rf}, \ref{fig:beta} and \ref{fig:fratio}), the optical data points were obtained from Figure 5 in \citet{ferrero2010}. The early time optical peak at $\sim 75\,\mathrm{s}$ after the trigger is not very prominent and dimmer than the optical afterglow peak at $\sim 500\,\mathrm{s}$ after the trigger (Figure \ref{fig:samplelcs}). The early time optical coverage is insufficient to perform the fit. If we take the first four optical data points and simply estimate the power-law indices we get roughly $\alpha _\mathrm{rise} \lesssim -1.1 $ and $\alpha _\mathrm{decay} \gtrsim 2.3$.
\par Optical data in $\mathrm{R_C}$ equivalent band (clear filter) were obtained from \citet{ferrero2010}.

\subsection{GRB~080928}
\label{sect:a080928}
\par \textbf{Literature}: No particular model for the early time optical peak was proposed in the literature, except that it is most likely produced via synchrotron radiation by the internal shocks \citep{rossi2011}. At later times the bumpy shape of the optical LC could be due to the energy injections into the forward shock \citep{rossi2011}.
\par \textbf{Energetics}: At redshift $z=1.692$ this GRB had an isotropic gamma-ray energy release of $E_{\gamma,\mathrm{iso}} = (1.44 \pm 0.92) \times 10^{52} \,\mathrm{ergs}$ in $1\,\mathrm{keV} - 10 \,\mathrm{MeV}$ band \citep{rossi2011}.
\par This GRB was also detected by the \textit{Fermi} GBM \citep{paciesas2008}. The detected emission is approximately coincident with the main emission detected by the \textit{Swift} BAT. Preliminary results of the spectral analysis of the GBM data show that the time-averaged spectrum can be fitted with the power-law function with power-law index $\beta = -1.80 \pm 0.08$ and fluence ($50-300\,\mathrm{keV}$) $f = (1.5 \pm 0.1)\times 10^{-6}\,\mathrm{erg/cm^2}$ \citep{paciesas2008}.
\par \textbf{Light curve}: There was a short lasting initial optical peak at $\sim 250\,\mathrm{s}$ after the trigger, which was detected by the \textit{Swift} UVOT in White filter \citep{rossi2011}. At later times the optical LC gets brighter, but stays complex and bumpy up to $\sim 10\,\mathrm{ks}$ after the trigger. After that it starts to decay with the power-law index $\alpha _\mathrm{decay} \approx 2.2$ \citep{rossi2011}.
\par The early time optical coverage is insufficient to examine the first optical peak and perform the fit (Figure \ref{fig:samplelcs}). If we take the first four optical data points and simply estimate the power-law indices we get roughly $\alpha _\mathrm{rise} \sim -3.0$ and $\alpha _\mathrm{decay} \sim 9.8$.
\par Optical data in the UVOT white band were obtained from \citet{rossi2011}. For the purpose of the spectral analysis we corrected the magnitudes for additional host galaxy absorption of $A_\mathrm{V}^\mathrm{rest} = 0.12$, or, assuming the MW-like extinction, for $A_\mathrm{w} = 0.36 \,\mathrm{mag}$ in the observer frame \citep{rossi2011}.

\subsection{GRB~081008}
\label{sect:a081008}
\par \textbf{Literature}: The first optical peak is interpreted as the onset of the afterglow emission, associated with the early time prompt phase \citep{yuan2010}. The second optical peak, which is superimposed on the early time optical afterglow and which is coincident with the gamma-ray activity, is probably from the late time internal shocks due to the central engine energy injection \citep{yuan2010}. However, this peak is not very prominent and it is thus not easy to study the origin in much detail (Figure \ref{fig:samplelcs}). Due to the complex optical LC it is possible that some internal shocks contribution is also present during the onset of the first optical peak, which could then be interpreted as the combination of the afterglow emission and the prompt optical emission.
\par \textbf{Energetics}: At redshift $z=1.967$ this GRB had an isotropic gamma-ray energy release of $E_{\gamma,\mathrm{iso}} \sim 6.3 \times 10^{52} \,\mathrm{ergs}$ in $0.1\,\mathrm{keV} - 10 \,\mathrm{MeV}$ band \citep{yuan2010}
\par \textbf{Light curve}: The results are preliminary, the optical data points were obtained from Figure 3 in \citet{yuan2010}. By fitting the sum of two contributions from Eq. (\ref{eq:beuermann}) to the early time optical LC in the time interval $42-757\,\mathrm{s}$ after the trigger, we get $\alpha _\mathrm{rise} = -5.7 \pm 1.6$ and $\alpha _\mathrm{decay} = 0.8 \pm 0.1$ for the first peak ($\chi _\mathrm{red} ^2$ is $1.1$). At later times the overall optical emission decays with power-law index $\alpha _\mathrm{decay} = 0.8 \pm 0.2$, consistent with the typical forward-external shock afterglow model \citep{yuan2010}.
\par Optical data in $r'$ equivalent band were obtained from \citet{yuan2010}. For the purpose of the spectral analysis we corrected the magnitudes for additional host galaxy absorption of $A_\mathrm{V}^\mathrm{rest} = 0.46$, or, assuming the MW-like extinction, for $A_\mathrm{r'} = 1.37 \,\mathrm{mag}$ in the observer frame \citep{yuan2010}.

\subsection{GRB~100901A}
\label{sect:a100901a}
\par \textbf{Literature}: The early time optical peak detected by the \textit{Swift} UVOT occurs simultaneously with the gamma-ray emission \citep{gomboc2013}. This suggests a common production site, i.e., the optical emission is the low energy tail of the prompt GRB emission, coming from the same region as high energy photons \citep{gorbovskoy2012}. This picture is consistent with the internal shock scenario, in which the correlation between the optical and gamma-ray emission is achieved via synchrotron emission from the same region.
\par \textbf{Energetics}: At redshift $z=1.408$ this GRB had an isotropic gamma-ray energy release of $E_{\gamma,\mathrm{iso}} = 6.3 \times 10^{52} \,\mathrm{ergs}$ in $1\,\mathrm{keV}-10\,\mathrm{MeV}$ band \citep{gorbovskoy2012}
\par \textbf{Light curve}: The results are preliminary (marked with $^\ast$ in Figures \ref{fig:lc_rf}, \ref{fig:beta} and \ref{fig:fratio}), the optical data points were obtained from \citet{pritchard2010} and \citet{gomboc2013}. By fitting the Beuermann profile (Eq. \ref{eq:beuermann}) to the early time optical peak in the time interval $305-545\,\mathrm{s}$ after the trigger, we get $\alpha _\mathrm{rise} = -11.2 \pm 4.4$ and $\alpha _\mathrm{decay} = 7.9 \pm 2.6$ ($\chi _\mathrm{red} ^2$ is $2.1$ with $4$ d.o.f.). At early times the LC is quite complex and probably composed of multiple contributions (the prompt optical emission and the onset of the external shock afterglow). At $\sim 0.3\,\mathrm{days}$ a powerful re-brightening occurs \citep{gomboc2013}.
\par Optical data in the UVOT u band were obtained from \citet{pritchard2010} and \citet{gomboc2013}. For the purpose of the spectral analysis we corrected the magnitudes for additional host galaxy absorption of $A_\mathrm{V}^\mathrm{rest} = 0.19$, or, assuming the SMC-like extinction, for $A_\mathrm{u} = 0.88 \,\mathrm{mag}$ in the observer frame \citep{gomboc2013}.

\subsection{GRB~100906A}
\label{sect:a100906a}
\par \textbf{Literature}: An optical peak occurs at $\approx 124\,\mathrm{s}$ after the trigger, during the gamma-ray emission and simultaneously with the gamma-ray peak \citep{gorbovskoy2012}. The smooth behaviour is indicative of the external-forward shock afterglow onset.
\par \textbf{Energetics}: At redshift $z=1.727$ this GRB had an isotropic gamma-ray energy release of $E_{\gamma,\mathrm{iso}} = (2.2 \pm 0.4) \times 10^{53} \,\mathrm{ergs}$ in $20\,\mathrm{keV}- 2\,\mathrm{MeV}$ band \citep{gorbovskoy2012}
\par \textbf{Light curve}: By fitting the Beuermann profile (Eq. \ref{eq:beuermann}) to the optical LC in the time interval $43-911\,\mathrm{s}$ after the trigger, we get $\alpha _\mathrm{rise} = -2.9 \pm 0.5$ and $\alpha _\mathrm{decay} = 1.0 \pm 0.1$ ($\chi _\mathrm{red} ^2$ is $0.9$ with $7$ d.o.f.). LC shows smooth and non-steep behaviour during the whole time (up until $10^4\,\mathrm{s}$ after the GRB trigger), consistent with the standard afterglow emission. No prompt optical signature is clearly present in the early time optical LC.
\par Optical data in V equivalent band were obtained from \citet{gorbovskoy2012}. 

\subsection{GRB~110205A}
\label{sect:a110205a}
\par \textbf{Literature}: The SED with two spectral breaks from the simultaneous optical and gamma-ray data at early times is consistent with the synchrotron emission in the fast cooling regime \citep{zheng2012}. This is most likely achieved via two-shell collisions in the internal shock model. Optical emission is thus the low energy tail of the prompt emission \citep{gao2011}. 
\par \textbf{Energetics}: At redshift $z=2.22$ this GRB had an isotropic gamma-ray energy release of $E_{\gamma,\mathrm{iso}} = 4.6^{+0.4}_{-0.7} \times 10^{53} \,\mathrm{ergs}$ in $1\,\mathrm{keV} - 10\,\mathrm{MeV}$ band \citep{zheng2012}
\par This GRB was also detected by the Konus-Wind \citep{golenetskii2011}. Results of the spectral analysis of the Konus-Wind data show that the spectrum is best fitted with the power-law function with exponential high-energy cut-off. Values of the parameters of the fit are: power-law index $\beta = -1.52 \pm 0.14$, cut-off energy $E_\mathrm{peak} = 222 \pm 74\,\mathrm{keV}$ and fluence ($20-1200\,\mathrm{keV}$) $f = (3.66 \pm 0.35)\times 10^{-5}\,\mathrm{erg/cm^2}$ \citep{golenetskii2011}.
\par \textbf{Light curve}: The early time optical LC is bumpy and its variability implies the internal shock origin. By fitting the Beuermann profile (Eq. \ref{eq:beuermann}) to the early time optical peak in the time interval $140-376\,\mathrm{s}$ after the trigger, we get $\alpha _\mathrm{rise} = -7.6 \pm 3.4$ and $\alpha _\mathrm{decay} = 3.4 \pm 0.9$ ($\chi _\mathrm{red} ^2$ is $1.1$ with $3$ d.o.f.). At later times, the optical LC shows the afterglow contribution from the reverse-external and forward-external shocks \citep{zheng2012, gao2011, gendre2012}.
\par Optical data in R band were obtained from \citet{gendre2012}. For the purpose of the spectral analysis we corrected the magnitudes for additional host galaxy absorption of $A_\mathrm{V}^\mathrm{rest} = 0.20$, or, assuming the SMC-like extinction, for $A_\mathrm{R} = 0.58 \,\mathrm{mag}$ in the observer frame \citep{gendre2012}.

\end{appendix}

\newpage

\input{phot_090727.tex}

\input{flux_beta_sample.tex}

\end{document}

%% file: phot_090727.tex
\begin{deluxetable}{cccccc}
\tablecaption{GRB~090727: Photometry.}
\tabletypesize{\scriptsize}
\tablehead{
\colhead{$t_\mathrm{mid}$ [s]} & 
\colhead{Exp [s]} &
\colhead{Telescope} & 
\colhead{Filter} &
\colhead{Magnitude} &
\colhead{$F_\nu^\mathrm{OPT}$ [mJy]} 
}
\startdata
$165.8$ & $10$ & LT & \textit{r'} & $19.66 \pm 0.28$ & $0.0514 \pm 0.0127$ \\ 
$186.8$ & $10$ & LT & \textit{r'} & $19.16 \pm 0.18$ & $0.08 \pm 0.0131$ \\ 
$209$ & $10$ & LT & \textit{r'} & $18.35 \pm 0.13$ & $0.1676 \pm 0.02$ \\ 
$287$ & $10$ & LT & \textit{r'} & $18.98 \pm 0.15$ & $0.094 \pm 0.0129$ \\ 
$308$ & $10$ & LT & \textit{r'} & $18.89 \pm 0.14$ & $0.1021 \pm 0.0131$ \\ 
$330.2$ & $10$ & LT & \textit{r'} & $19.84 \pm 0.34$ & $0.0443 \pm 0.0134$ \\ 
$472.2$ & $120$ & LT & \textit{r'} & $20.22 \pm 0.15$ & $0.03 \pm 0.0041$ \\ 
$985.2$ & $120$ & LT & \textit{r'} & $20.11 \pm 0.14$ & $0.0332 \pm 0.0043$ \\ 
$1117.2$ & $120$ & LT & \textit{r'} & $20.32 \pm 0.14$ & $0.0273 \pm 0.0035$ \\ 
$2038.2$ & $360$ & LT & \textit{r'} & $20.38 \pm 0.14$ & $0.0259 \pm 0.0033$ \\ 
$3457.2$ & $360$ & LT & \textit{r'} & $20.48 \pm 0.14$ & $0.0236 \pm 0.003$ \\ 
$4443.6$ & $60$ & LT & \textit{r'} & $20.78 \pm 0.3$ & $0.0184 \pm 0.005$ \\ 
$4687.2$ & $120$ & LT & \textit{r'} & $20.7 \pm 0.14$ & $0.0193 \pm 0.0025$ \\ 
$5268$ & $240$ & LT & \textit{r'} & $20.7 \pm 0.14$ & $0.0193 \pm 0.0025$ \\ 
$6249$ & $360$ & LT & \textit{r'} & $20.87 \pm 0.14$ & $0.0165 \pm 0.0021$ \\ 
$7650$ & $360$ & LT & \textit{r'} & $21.24 \pm 0.14$ & $0.0117 \pm 0.0015$ \\ 
$18882.6$ & $1800$ & LT & \textit{r'} & $21.67 \pm 0.1$ & $0.0079 \pm 0.0007$ \\ 
$42146.4$ & $1800$ & FTN & \textit{R} & $22.65 \pm 0.33$ & $0.0033 \pm 0.001$ \\ 
$52320$ & $1800$ & FTN & \textit{R} & $22.72 \pm 0.42$ & $0.0032 \pm 0.0012$ \\
$195044$ & $1800$ & LT & \textit{r'} & $>23.32$ & $<0.0017$ \\
 & & & & \\
$643.8$ & $120$ & LT & \textit{i'} & $20.34 \pm 0.18$ & $0.027 \pm 0.0044$ \\ 
$1354.8$ & $240$ & LT & \textit{i'} & $19.79 \pm 0.1$ & $0.0444 \pm 0.0041$ \\ 
$2512.2$ & $360$ & LT & \textit{i'} & $19.66 \pm 0.1$ & $0.05 \pm 0.0046$ \\ 
$4858.2$ & $120$ & LT & \textit{i'} & $20.11 \pm 0.14$ & $0.0332 \pm 0.0043$ \\ 
$5578.8$ & $240$ & LT & \textit{i'} & $20.13 \pm 0.11$ & $0.0325 \pm 0.0033$ \\ 
$6720$ & $360$ & LT & \textit{i'} & $20.28 \pm 0.11$ & $0.0283 \pm 0.0029$ \\ 
$20868.6$ & $1800$ & LT & \textit{i'} & $21.03 \pm 0.14$ & $0.0142 \pm 0.0018$ \\ 
$44539.8$ & $1500$ & FTN & \textit{i'} & $22.02 \pm 0.35$ & $0.006 \pm 0.0019$ \\ 
$122887$ & $1800$ & FTN & \textit{i'} & $22.13 \pm 0.39$ & $0.0055 \pm 0.0019$ \\ 
 & & & & \\
$820.2$ & $120$ & LT & \textit{z'} & $19.48 \pm 0.24$ & $0.0601 \pm 0.0131$ \\ 
$1666.8$ & $240$ & LT & \textit{z'} & $19.16 \pm 0.24$ & $0.0807 \pm 0.0175$ \\ 
$2982$ & $360$ & LT & \textit{z'} & $19.46 \pm 0.24$ & $0.0612 \pm 0.0133$ \\ 
$5539.2$ & $360$ & LT & \textit{z'} & $19.55 \pm 0.24$ & $0.0563 \pm 0.0123$ \\ 
$7183.8$ & $360$ & LT & \textit{z'} & $20.05 \pm 0.25$ & $0.0356 \pm 0.0081$
\enddata
\tablecomments{\label{tab:phot090727}Magnitudes are corrected for the Galactic extinction.}
\end{deluxetable}

%% file: flux_beta_sample.tex
\LongTables
\begin{deluxetable*}{ccccccccc}
\tablecaption{Early time optical and gamma-ray flux densities and spectral indices.}
\tabletypesize{\scriptsize}
\tablewidth{\linewidth}
\tablehead{
\colhead{GRB} & 
\colhead{$t_\mathrm{mid}$} & 
\colhead{Exp} & 
\colhead{$\nu_\mathrm{OPT}$} &
\colhead{$F_\nu^\mathrm{OPT}$} &
\colhead{$\nu_\gamma$} &
\colhead{$F_\nu^\gamma$} &
\colhead{$\beta _\gamma$} &
\colhead{$\beta _\mathrm{OPT-\gamma}$} 
\\
\colhead{} & 
\colhead{[s]} & 
\colhead{[s]} & 
\colhead{[$10^{14}\,\mathrm{Hz}$]} & 
\colhead{[$\mathrm{mJy}$]} & 
\colhead{[$10^{18}\,\mathrm{Hz}$]} & 
\colhead{[$\mathrm{\mu Jy}$]} & 
\colhead{} & 
\colhead{}
} 
\startdata
990123 & $24.7$ & $5$ & $5.5$ & $78.2 \pm 5.03$ & $24$ & $4450 \pm 0$ & $-0.4 \pm 0.01$ & $0.27 \pm 0.01$ \\ 
 & $49.9$ & $5$ & $5.5$ & $1067.5 \pm 19.66$ & $24$ & $1630 \pm 0$ & $-0.4 \pm 0.01$ & $0.61 \pm 0$ \\ 
 & $75.2$ & $5$ & $5.5$ & $384.11 \pm 10.61$ & $24$ & $1710 \pm 0$ & $-0.4 \pm 0.01$ & $0.51 \pm 0$ \\ 
041219A & $239.2$ & $72.5$ & $4.8$ & $3.03 \pm 0.28$ & $5$ & $734 \pm 39$ & $0.15 \pm 0.15$ & $0.15 \pm 0.01$ \\ 
 & $303$ & $30$ & $4.8$ & $10.04 \pm 0.92$ & $5$ & $3600 \pm 190$ & $0.51 \pm 0.03$ & $0.11 \pm 0.01$ \\ 
 & $367.7$ & $70.5$ & $4.8$ & $3.65 \pm 0.33$ & $5$ & $2882 \pm 75$ & $0.74 \pm 0.02$ & $0.03 \pm 0.01$ \\ 
 & $494.3$ & $157.7$ & $4.8$ & $1.12 \pm 0.2$ & $5$ & $583 \pm 31$ & $1.34 \pm 0.09$ & $0.07 \pm 0.02$ \\  
050820A & $267.4$ & $30$ & $4.8$ & $2.55 \pm 0.05$ & $25$ & $453 \pm 17$ & $0.37 \pm 0.06$ & $0.16 \pm 0$ \\ 
 & $426.8$ & $30$ & $4.8$ & $4.73 \pm 0.09$ & $25$ & $314 \pm 16$ & $0.42 \pm 0.08$ & $0.25 \pm 0$ \\  
 & $530.7$ & $30$ & $4.8$ & $4.35 \pm 0.08$ & $27$ & $138 \pm 15$ & $0.71 \pm 0.14$ & $0.32 \pm 0.01$ \\ 
060526$^\ast$ & $246.3$ & $10$ & $5.5$ & $1.17 \pm 0.21$ & $6$ & $227.3 \pm 41.1$ & $1.07 \pm 0.18$ & $0.18 \pm 0.03$ \\ 
 & $256.3$ & $10$ & $5.5$ & $2.36 \pm 0.3$ & $6$ & $151.7 \pm 39$ & $0.91 \pm 0.24$ & $0.29 \pm 0.03$ \\ 
060729$^\ast$ & $81.1$ & $5$ & $4.8$ & $0.62 \pm 0.14$ & $7$ & $327.6 \pm 79.4$ & $0.63 \pm 0.15$ & $0.07 \pm 0.03$ \\ 
 & $95.6$ & $5$ & $4.8$ & $1.84 \pm 0.24$ & $7.5$ & $298.6 \pm 55.6$ & $1 \pm 0.15$ & $0.19 \pm 0.02$ \\ 
 & $117.5$ & $5$ & $4.8$ & $0.6 \pm 0.2$ & $7.5$ & $61.1 \pm 51.1$ & $0.97 \pm 0.62$ & $0.24 \pm 0.09$ \\ 
060904B$^\ast$ & $29.9$ & $5$ & $4.8$ & $0.58 \pm 0.2$ & $5.1$ & $23 \pm 32.7$ & $0.52 \pm 0.91$ & $0.35 \pm 0.16$ \\ 
 & $157.3$ & $20$ & $4.8$ & $0.62 \pm 0.12$ & $4.6$ & $136.8 \pm 31.9$ & $1.42 \pm 0.35$ & $0.17 \pm 0.03$ \\ 
061007$^\ast$ & $43.5$ & $5$ & $4.8$ & $688.78 \pm 12.69$ & $6.5$ & $1716.5 \pm 117.3$ & $-0.06 \pm 0.04$ & $0.63 \pm 0.01$ \\ 
 & $57.9$ & $5$ & $4.8$ & $1164.34 \pm 21.45$ & $5.1$ & $2672 \pm 226.4$ & $0 \pm 0.03$ & $0.66 \pm 0.01$ \\ 
 & $80.3$ & $5$ & $4.8$ & $995.46 \pm 9.17$ & $6.5$ & $375 \pm 61.1$ & $0.69 \pm 0.12$ & $0.83 \pm 0.02$ \\ 
 & $94.5$ & $5$ & $4.8$ & $1219.22 \pm 22.46$ & $7$ & $161.6 \pm 53.9$ & $0.74 \pm 0.25$ & $0.93 \pm 0.03$ \\ 
061121$^\ast$ & $64.4$ & $2$ & $5.5$ & $1.57 \pm 0.34$ & $5.1$ & $1562.8 \pm 172.6$ & $0.58 \pm 0.07$ & $0 \pm 0.03$ \\ 
 & $68.4$ & $2$ & $5.5$ & $2.99 \pm 0.48$ & $5.6$ & $3325.4 \pm 244.9$ & $0.27 \pm 0.04$ & $-0.01 \pm 0.02$ \\ 
 & $72.4$ & $2$ & $5.5$ & $3.49 \pm 0.5$ & $5.6$ & $4830.8 \pm 326.9$ & $0.51 \pm 0.04$ & $-0.04 \pm 0.02$ \\ 
 & $76.4$ & $2$ & $5.5$ & $7.32 \pm 1$ & $5.1$ & $2966.2 \pm 273.2$ & $0.36 \pm 0.04$ & $0.1 \pm 0.02$ \\ 
 & $80.4$ & $2$ & $5.5$ & $4.82 \pm 0.66$ & $5.6$ & $646.6 \pm 102.6$ & $0.74 \pm 0.13$ & $0.23 \pm 0.02$ \\ 
 & $84.4$ & $2$ & $5.5$ & $2.7 \pm 0.42$ & $5.6$ & $307.9 \pm 90.9$ & $0.62 \pm 0.21$ & $0.25 \pm 0.04$ \\ 
080310$^\ast$ & $298$ & $30$ & $5.5$ & $1.17 \pm 0.08$ & $4.6$ & $144.3 \pm 47.3$ & $1.52 \pm 0.53$ & $0.23 \pm 0.04$ \\ 
080319B$^\ast$ & $10.2$ & $1.3$ & $5.5$ & $6960.42 \pm 1143.48$ & $8$ & $5424.6 \pm 324.8$ & $0.12 \pm 0.04$ & $0.75 \pm 0.02$ \\ 
 & $18.3$ & $1.3$ & $5.5$ & $31404.3 \pm 1156.45$ & $6$ & $6829 \pm 363.7$ & $0.07 \pm 0.03$ & $0.91 \pm 0.01$ \\ 
 & $32.8$ & $1.3$ & $5.5$ & $17578.8 \pm 647.33$ & $5.1$ & $4749.4 \pm 420.2$ & $0.01 \pm 0.04$ & $0.9 \pm 0.01$ \\ 
 & $35.4$ & $1.3$ & $5.5$ & $26844.8 \pm 741.56$ & $6$ & $5323.1 \pm 298.7$ & $0.19 \pm 0.04$ & $0.92 \pm 0.01$ \\ 
 & $39.3$ & $1.3$ & $5.5$ & $20552.29 \pm 567.74$ & $5.1$ & $5153.3 \pm 452.7$ & $0.04 \pm 0.04$ & $0.91 \pm 0.01$ \\ 
 & $43.2$ & $1.3$ & $5.5$ & $30252.98 \pm 557.22$ & $7.5$ & $4487.5 \pm 281.8$ & $0.22 \pm 0.04$ & $0.93 \pm 0.01$ \\ 
 & $51$ & $1.3$ & $5.5$ & $17099.72 \pm 629.69$ & $7.5$ & $3423.4 \pm 240.9$ & $0.83 \pm 0.05$ & $0.89 \pm 0.01$ \\ 
 & $61.4$ & $1.3$ & $5.5$ & $6116.94 \pm 561.8$ & $6$ & $624.1 \pm 119.8$ & $0.94 \pm 0.18$ & $0.99 \pm 0.02$ \\ 
080810 & $38$ & $5$ & $4.8$ & $10.71 \pm 0.3$ & $4.6$ & $106.2 \pm 62.5$ & $0.52 \pm 0.66$ & $0.5 \pm 0.06$ \\ 
 & $52$ & $5$ & $4.8$ & $20.41 \pm 0.56$ & $7$ & $283.2 \pm 61.2$ & $0.64 \pm 0.15$ & $0.45 \pm 0.02$ \\ 
 & $67$ & $5$ & $4.8$ & $24.76 \pm 0.68$ & $5.1$ & $78.2 \pm 60.3$ & $0.5 \pm 0.52$ & $0.62 \pm 0.08$ \\ 
 & $95$ & $5$ & $4.8$ & $24.31 \pm 0.67$ & $7$ & $54.4 \pm 41.5$ & $0.96 \pm 0.94$ & $0.64 \pm 0.08$ \\ 
080905B & $60.2$ & $10$ & $4.8$ & $2.01 \pm 0.35$ & $5.6$ & $102.2 \pm 44$ & $0.5 \pm 0.27$ & $0.32 \pm 0.05$ \\ 
 & $75.9$ & $10$ & $4.8$ & $2.45 \pm 0.42$ & $5.1$ & $115.7 \pm 39.1$ & $0.77 \pm 0.32$ & $0.33 \pm 0.04$ \\ 
 & $87.2$ & $10$ & $4.8$ & $1.84 \pm 0.37$ & $5.6$ & $104.3 \pm 38.6$ & $0.37 \pm 0.24$ & $0.31 \pm 0.05$ \\ 
080928$^\ast$ & $209$ & $20$ & $5.5$ & $0.08 \pm 0.02$ & $5.1$ & $332.1 \pm 39.3$ & $0.8 \pm 0.09$ & $-0.16 \pm 0.03$ \\ 
081008$^\ast$ & $60.1$ & $5$ & $4.8$ & $33.98 \pm 1.1$ & $9$ & $78.7 \pm 53.1$ & $0.8 \pm 0.39$ & $0.62 \pm 0.07$ \\ 
 & $68.6$ & $5$ & $4.8$ & $43.15 \pm 1.64$ & $5.6$ & $64.5 \pm 53.6$ & $0.86 \pm 1.05$ & $0.7 \pm 0.09$ \\ 
 & $99.6$ & $5$ & $4.8$ & $51.41 \pm 1.99$ & $5.1$ & $239.8 \pm 61.1$ & $0.99 \pm 0.25$ & $0.58 \pm 0.03$ \\ 
 & $107.8$ & $5$ & $4.8$ & $54.8 \pm 1.79$ & $5.6$ & $294.4 \pm 61.9$ & $1.16 \pm 0.21$ & $0.56 \pm 0.02$ \\ 
 & $115.7$ & $5$ & $4.8$ & $50.92 \pm 1.29$ & $7.5$ & $257.7 \pm 57.5$ & $0.66 \pm 0.14$ & $0.55 \pm 0.02$ \\ 
 & $142.7$ & $20$ & $4.8$ & $44.84 \pm 0.85$ & $7$ & $55 \pm 26.2$ & $1.19 \pm 0.51$ & $0.7 \pm 0.05$ \\ 
 & $171.7$ & $20$ & $4.8$ & $38.17 \pm 1.29$ & $4.6$ & $31.3 \pm 29.7$ & $1.22 \pm 1.75$ & $0.78 \pm 0.1$ \\ 
090727 & $209$ & $10$ & $4.8$ & $0.17 \pm 0.02$ & $8.5$ & $21.8 \pm 27.8$ & $0.5 \pm 1.05$ & $0.21 \pm 0.13$ \\ 
 & $287$ & $10$ & $4.8$ & $0.09 \pm 0.01$ & $6$ & $32.7 \pm 30.5$ & $1.41 \pm 2.87$ & $0.11 \pm 0.1$ \\ 
100901A$^\ast$ & $380$ & $30$ & $8.6$ & $0.56 \pm 0.1$ & $5.1$ & $64.1 \pm 22.2$ & $0.34 \pm 0.23$ & $0.25 \pm 0.04$ \\ 
 & $410$ & $30$ & $8.6$ & $0.97 \pm 0.12$ & $7.5$ & $47.4 \pm 21.5$ & $0.3 \pm 0.29$ & $0.33 \pm 0.05$ \\ 
100906A & $48.6$ & $10$ & $5.5$ & $8.26 \pm 0.3$ & $4.6$ & $370.4 \pm 59.4$ & $1.17 \pm 0.18$ & $0.34 \pm 0.02$ \\ 
 & $78.8$ & $10$ & $5.5$ & $38.82 \pm 1.07$ & $5.1$ & $234.2 \pm 44.8$ & $1.05 \pm 0.23$ & $0.56 \pm 0.02$ \\ 
 & $115.4$ & $20$ & $5.5$ & $46.67 \pm 1.29$ & $4.6$ & $727.8 \pm 84.5$ & $1.75 \pm 0.12$ & $0.46 \pm 0.01$ \\ 
110205A$^\ast$ & $113$ & $6$ & $4.8$ & $0.35 \pm 0.15$ & $7$ & $342.9 \pm 55.1$ & $0.75 \pm 0.12$ & $0 \pm 0.05$ \\ 
 & $137$ & $6$ & $4.8$ & $0.5 \pm 0.16$ & $7$ & $532.7 \pm 57.2$ & $0.71 \pm 0.08$ & $-0.01 \pm 0.03$ \\ 
 & $181$ & $30$ & $4.8$ & $0.39 \pm 0.03$ & $6.5$ & $362.7 \pm 28$ & $0.86 \pm 0.06$ & $0.01 \pm 0.01$ \\ 
 & $226$ & $30$ & $4.8$ & $0.93 \pm 0.07$ & $7$ & $425.6 \pm 29.1$ & $0.7 \pm 0.05$ & $0.08 \pm 0.01$ \\ 
 & $271$ & $30$ & $4.8$ & $0.48 \pm 0.04$ & $5.6$ & $170.4 \pm 24.2$ & $0.84 \pm 0.12$ & $0.11 \pm 0.02$ 
\enddata
\tablecomments{Optical flux densities are corrected for the Galactic extinction using \citet{schlegel1998} and for GRBs marked with $^\ast$ flux densities are also corrected for the host galaxy extinction (references are given in Appendices \ref{sect:a990123} - \ref{sect:a110205a}). $t_\mathrm{mid}$ is given since the GRB trigger time. If the gamma-ray emission during the exposure time of the optical emission was too weak to obtain good gamma-ray spectrum and $\beta_\gamma$, we extended the gamma-ray interval to twice the exposure time in the optical. The confidence intervals for $\beta_\gamma$ and $\beta_\mathrm{OPT-\gamma}$ are at $90\%$. \\
Optical emission references are given in Appendices \ref{sect:a990123}-\ref{sect:a110205a}. Gamma-ray flux densities and spectral indices were extracted from the analysis of the \textit{Swift} BAT data (Section \ref{sect:specprop}), except for GRB~990123, GRB~041219A and GRB~050820A, for which $F_\nu^\gamma$ and $\beta_\gamma$ were taken from \citet{yost2007a} and \citet{yost2007b}.}
\label{tab:photomspec}
\end{deluxetable*}